\documentclass{aastex}
\usepackage{emulateapj5}

\newcommand{\kms}{\,km~s$^{-1}$}
\newcommand{\sqcm}{\,cm$^{-2}$}  
\newcommand{\fuse}{\emph{FUSE}} 
\newcommand{\hst}{\emph{HST}}
\newcommand{\os}{\ion{O}{6}}
\newcommand{\nf}{\ion{N}{5}}
\newcommand{\cf}{\ion{C}{4}}
\newcommand{\sif}{\ion{Si}{4}}
\newcommand{\hi}{\ion{H}{1}}
\newcommand{\ha}{H$\alpha$}
\newcommand{\pga}{PG~1259+593}
\newcommand{\pgb}{PG~1351+640}

\begin{document}

\shorttitle{Highly Ionized Gas Surrounding Complex C}
\shortauthors{Fox et al.}

\title{Highly Ionized Gas Surrounding High Velocity Cloud Complex
  C\footnotemark[1]}
\footnotetext[1]{Based on observations from the NASA-CNES-CSA {\it Far
  Ultraviolet Spectroscopic Explorer} mission, operated by Johns
  Hopkins University, supported by NASA contract NAS 5-32985, and from the
  NASA/ESA {\it Hubble Space Telescope}, obtained at the Space Telescope
  Science Institute, which is operated by the Association of
  Universities for Research in Astronomy, Inc., under NASA contract
  NAS 5-26555.} 
\author{Andrew J. Fox\altaffilmark{2}, Blair D. Savage\altaffilmark{2}, 
Bart P. Wakker\altaffilmark{2}, Philipp Richter\altaffilmark{3}, 
Kenneth R. Sembach\altaffilmark{4}, \& Todd M. Tripp\altaffilmark{5, 6}}

\altaffiltext{2}{Department of Astronomy, University of Wisconsin -
Madison, 475 North Charter St., Madison, WI 53706} 
\altaffiltext{3}{Osservatorio Astrofisico di Arcetri, Largo E. Fermi, 5,
I-50125, Florence, Italy}
\altaffiltext{4}{Space Telescope Science Institute, 3700 San Martin
Drive, Baltimore, MD 21218} 
\altaffiltext{5}{Princeton University Observatory, Peyton Hall, Ivy
Lane, Princeton, NJ 08544} 
\altaffiltext{6}{Department of Astronomy, University of Massachusetts, 
Amherst, MA 01003}

\begin{abstract}
We present \emph{Far Ultraviolet Spectroscopic Explorer}
and \emph{Hubble Space Telescope} observations of high, intermediate,
and low ion absorption  
in high-velocity cloud Complex C along the lines of sight toward five 
active galaxies. Our purpose is to investigate the idea that
Complex C is surrounded by an envelope of highly ionized
material, arising from the interaction between the cloud and a hot
surrounding medium. We measure column densities of high-velocity high
ion absorption, and compare the kinematics of low, intermediate, and
high ionization gas along the five sight lines. We find that in all
five cases, the \hi\ and \os\ high-velocity components are centered
within 20\kms\ of one another, with an average displacement of
$<\bar{v}_{\mathrm{O~VI}}-\bar{v}_{\mathrm{H~I}}>=3\pm12$\kms. 
In those directions where the \hi\ emission extends to more negative
velocities (the so-called high-velocity ridge), so does the \os\ absorption.
The kinematics of \ion{Si}{2} are also similar to those of \os, with 
$<\bar{v}_{\mathrm{O~VI}}-\bar{v}_{\mathrm{Si~II}}>=0\pm15$\kms. 
We compare our high ion column density ratios to the predictions of
various models, adjusted to account for both recent updates to the solar
elemental abundances, and for the relative elemental abundance
ratios in Complex~C. Along the \pga\ sight line, we measure 
$N$(\sif)/$N$(\os) $=0.10\pm0.02$, $N$(\cf)/$N$(\os) $=0.35^{+0.05}_{-0.06}$,
and $N$(\nf)/$N$(\os) $<0.07$ (3$\sigma$). These ratios are
inconsistent with collisional 
ionization equilibrium at one kinetic temperature. Photoionization by
the extragalactic background is ruled out as the source of the 
high ions since the path lengths required would make HVCs unreasonably 
large; photoionization by radiation from the disk of the Galaxy
also appears unlikely since the emerging photons are not energetic enough
to produce \os.
By themselves, ionic ratios are insufficient to discriminate between
various ionization models, but by considering the absorption 
kinematics as well we consider the most likely origin for the
highly ionized high-velocity gas to be at the conductive or turbulent
interfaces between
the neutral/warm ionized components of Complex~C and a surrounding hot
medium. The presence of interfaces on the surface
of HVCs provides indirect evidence for the existence of a hot medium
in which the HVCs are immersed. This medium could be a hot
($T\gtrsim10^6$\,K) extended Galactic corona or hot gas in the Local Group.
\end{abstract}

\section{Introduction}

Complex~C is an extensive high-velocity cloud (HVC) covering 1700 square 
degrees in the northern Galactic sky and falling at an average LSR
velocity of $-130$\kms\ onto the Milky Way. Because of its low
metallicity ($\approx0.1-0.2$ solar), Complex~C is suspected of comprising 
either intergalactic gas or material tidally stripped from a nearby
galaxy, with perhaps
some contribution from upwelling Galactic outflow. HVCs are
defined as clouds whose observed radial velocities deviate
significantly with those expected from Galactic rotation,
corresponding 
in practice to clouds with $|v_{LSR}|>100$\kms. They have
traditionally been studied in neutral hydrogen 21\,cm emission, and
more recently with both \ha\ emission measurements and absorption line
studies. 

Until 1995, all the absorption 
detected in HVCs had been in lines of neutral or low ionization
species. Highly ionized gas was then discovered with
the detections of high-velocity \cf\ in the spectra of Mrk~509 and
PKS~2155--304 \citep{Se95,Se99}. Following the launch
of the \emph{Far Ultraviolet Spectroscopic Explorer} (\fuse) satellite in
the summer of 1999, \os\ has been repeatedly detected in absorption in
HVCs; the first high-velocity \os\ results were reported by \citet{Se00} and
\citet{Mu00}. More recently, a survey of high-velocity \os\ has been
completed by \citet*[][hereafter S03]{Se03a}. Since \os\ is an ion that exists
at temperatures of a few$\;\times10^5$ K \citep{SD93}, its presence in
HVCs poses a variety of
intriguing questions: primarily, what physical process produces the
\os? Does the ion originate at some form of interface between the
neutral and warm ionized components of HVCs and a surrounding hot
medium? 
Can photoionization by the hard extragalactic radiation field play an
important role? S03 began to address these issues with a study of 102
sight lines, 58 of which displayed high-velocity \os\ absorption. 

In this project we focus our attention on five extragalactic sight lines
(Mrk~279, Mrk~817, Mrk~876, \pga, and \pgb) passing
through HVC Complex~C, in order to determine the inter-relationships
between the various phases of gas present in HVCs. Our approach is to 
compare the kinematics and
column densities of high and low ion absorption in high-velocity  
gas. Clues concerning the physical conditions within the absorbing
material are provided by observing the extent and structure of the
absorption profiles in different species.

In Figure 1 we show a color map of Complex~C from the Dwingeloo HVC
survey \citep{HW88}, displaying the velocity structure within the 
complex. Gas covering a range in velocities of $\approx100$\kms\ is
present in Complex C. The observed range in column densities and
velocities indicates a complex system with a large amount of sub-structure. 
The sight lines studied in this paper are marked with stars.

Our paper is structured as follows: In \S2 we give a brief summary of
previous studies of Complex~C. Section 3 contains a description of our
observations and 
data reduction. We present spectroscopic measurements of high-velocity,
highly ionized absorption in \S4. In \S5 we display our spectra and
compare the high and low ion absorption along five Complex~C sight
lines. In \S6 we discuss and review various models that might explain
the existence of highly ionized gas in HVCs. In \S7 we discuss the
influence of non-solar abundance patterns on the model
predictions. The theories are 
compared with our observations in \S8, using ionic ratios and
kinematic information. We mention interesting results from sight lines
passing near to Complex~C in \S9. Our results are summarized in \S10. 

\section{Previous Studies of Complex~C}

Detailed abundances studies of Complex~C have been carried out by a
number of authors \citep*{Wa99, Ri01a, Gi01, Co03a, Tr03, Se03b}, who have all
used \hst\ and \fuse\ spectroscopy to determine column densities of various ionic
species through the complex, and combined them with \hi\ 21\,cm
emission measurements to derive elemental abundances. These studies
have found values of (\ion{O}{1}/\hi) between 0.09 and 0.25 times
solar, and no evidence for depletion onto dust. Precise measurements
are difficult because of a number of sources of error, particularly
uncertain ionization
corrections, the difficulty of measuring a precise $b$ value, and the
fact that \hi\ radio emission measurements sample 
a beam far larger than that probed in absorption measurements.

Relative elemental abundance ratios are important in understanding the
history of Complex~C. \citet{RD92} find that in the Magellanic Clouds
log\,(N/O)$_{LMC}=-1.21$ and log\,(N/O)$_{SMC}=-1.40$,
and \citet{KS96} found that log\,(N/O) typically lies 
between $-1.7$ and $-1.5$ in dwarf galaxies with
log\,$A_{\mathrm{O}}<-4.0$. An observed value in Complex~C is
log\,(N/O)$~=-1.7$ \citep[][towards \pga]{Co03a}, in line with these 
measurements in other locations\footnotemark[7]. 
\footnotetext[7]{Throughout this paper we use the conventions
$[\mathrm{X}/\mathrm{Y}]=$ log($N_\mathrm{X}/N_\mathrm{Y})-$log($N_\mathrm{X}/N_\mathrm{Y})_{\odot}$, $(\mathrm{X}/\mathrm{Y})=N_\mathrm{X}/N_\mathrm{Y}$,
$A_\mathrm{X}=(\mathrm{X}/\mathrm{H})$, and $A_\mathrm{X}^{\odot}=(\mathrm{X}/\mathrm{H})_{\odot}$.} 

A conclusive direct distance determination would be an invaluable piece of
information in studying Complex~C; unfortunately, such a determination
is difficult to make. The strongest constraint that exists
is a lower limit of 6.1\,kpc \citep{Wa01}, based upon the non-detection of
\ion{Ca}{2} in absorption towards stars at known distances lying in this part
of the sky. Note, however, that this non-detection is not strongly
significant \citep[as defined by][]{Wa01}. A more reliable lower limit to the
distance is 1.3\,kpc. 

It has been suggested that HVCs may represent material taking part in
a Galactic fountain \citep{SF76, Br80}. However, the low
metallicities observed make this scenario unlikely for Complex~C,
since fountain gas, having been blown out of the disk by Type II
supernovae, should have solar or super-solar rather than sub-solar 
abundances, and a solar N/O ratio of log\,(N/O)$_{\odot}=-0.8$. Substantial
dilution by primordial gas would be necessary to bring fountain gas
down to the observed level of metallicity. We note that the
approximately solar abundances measured in intermediate velocity
clouds \citep[IVCs;][]{Ri01a} suggest they are formed by Galactic
material; it is the IVCs that are consequently thought to be tracing
the fountain circulation. 

\section{Observations and Data Reduction}

The \fuse\ satellite \citep{Mo00} operates in the far-ultraviolet wavelength
regime between 905 and 1187\,\AA, providing medium resolution
($R\approx15,000$, corresponding to $\approx20$\kms\ FWHM) spectra
with enough sensitivity that faint extragalactic sources can be observed. 
The \fuse\ data analyzed in this paper were obtained from the 
Multimission Archive at the Space Telescope Science Institute
(MAST)\footnotemark[8].
\footnotetext[8]{Online at {\tt http://archive.stsci.edu/mast.html}}

In order to investigate the highly ionized gas in Complex~C, we took
all the sight lines through Complex~C observed with the \fuse\ satellite and
applied a set of selection criteria to identify those directions
containing  useful high ion information. These criteria took the form
of a series of constraints, that eventually left us with our sample. To be
classified as going through Complex~C, we required that a sight line
had to lie within the $2\times10^{18}$\sqcm\ contour of $N$(\hi) in
the \citet{HW88} Dwingeloo HVC survey. Constraint 1 eliminated sight
lines with poor signal-to-noise, i.e. S/N$~<5$ near 1030\,\AA\ in the
10-pixel-rebinned \fuse\ spectrum. Constraint 2 looked for sight
lines with good \os\ data, and rejected those with blending or
continuum placement problems. Finally, Constraint 3 looked for sight
lines that have been observed with either the Goddard High Resolution
Spectrograph (GHRS) or the Space Telescope Imaging Spectrograph (STIS)
on \hst, at better than 30\kms\ resolution. Those that
were not were eliminated. The reason for Constraint 3 is that with
\hst\ data we can form ionic ratios with \os, or at least upper
limits, to investigate the ionization mechanism. Such an approach
cannot be achieved with \fuse\ data alone. Applying these conditions generated
our sample of five sight lines, namely Mrk~279, Mrk~817, Mrk~876,
\pga, and \pgb, each of which have good \fuse\ and \hst\ data. Of our
chosen five sight lines, two (Mrk~279 and Mrk~817) have GHRS 
spectra of the region near 1240\,\AA, where the \nf\ doublet lies, and 
three (Mrk~876, \pga\ and \pgb) have STIS spectra covering this region.
In the case of \pga, we were also able to use STIS data of
\sif, \cf, and various low-ionization species. Consequently, this
remains the sight line for which we have the best quality data
available, and the strongest insight into ionization processes.

A summary of basic target data is given in Table 1, where for each
sight line we include Galactic longitude $l$, Galactic latitude
$b$, redshift of source $z$, type of source, apparent $B$
magnitude of source, Galactic foreground interstellar reddening
$E(B-V)$, high-velocity neutral hydrogen column density $N$(\hi),
heliocentric to LSR velocity correction $\Delta v_{LSR}$,  and recent
measurements of [O/H] and [N/H] in Complex~C by \citet{Co03a}.  

In the following two subsections we describe our data handling
procedures for our two separate datasets. 

\subsection{FUSE}

Details of our \fuse\ observations are given in Table 2. The data were
reduced using the data reduction pipeline {\tt CALFUSE} v2.1.6, except
for Mrk~279, for which we used v2.0.5. This anomaly was because of
compatibility problems between early \fuse\ 
datasets (such as the Mrk~279 observations) and the newer versions of the
{\tt CALFUSE} pipeline. The difference in velocity calibration caused by this
inconsistency is expected to be minimal, since all {\tt CALFUSE}
versions after 2.0 have greatly improved velocity scale accuracy over
earlier ones.  

Despite the improvements made in the calibration software, we still
found it necessary to apply velocity shifts to each spectrum to
register the velocity scale correctly. Our approach was to tie the
\ion{Si}{2} $\lambda1020.699$, \ion{C}{2} $\lambda1036.337$,
\ion{O}{1} $\lambda1039.230$, and \ion{Ar}{1} $\lambda1048.220$
absorption lines to the centers of emission observed in \hi, using 
Effelsberg 100\,m radio telescope 21\,cm observations. This process transforms
the velocity scale into the Local Standard of Rest (LSR) frame. After
using Gaussian component fitting software to measure the velocity centroids of
components in neutral gas absorption lines, we determined the velocity
correction necessary to align these components with the corresponding
components seen in
\hi. Where possible, we tied the scale to the high-velocity components
of \hi, since these velocities were often cleaner and more distinct,
and hence easier to measure. The actual lines used for alignment 
in each case depended on the individual spectrum. % (see below).
We checked for velocity scale differences between different channels and
different exposures within the same channel. The shifts we applied
were all in the range 10 to 50\kms. The Effelsberg 100\,m radio
telescope has a beam size of $9\arcmin\!.7$, and a velocity resolution of
$\approx1$\kms; the \hi\ observations we used are described in detail
in \citet{WK01}.

Our alignment procedure has the advantage that \hi\ spectra are extremely
reliable in their velocity calibration, since radio frequencies can be
measured with very high precision. Additionally, when observing an
\hi\ spectrum towards an extragalactic source, one has the advantage of
knowing that the source is beyond the emitting gas, so absorption
lines and emission lines can be reliably compared. Thus the only
potential source of errors in our alignment process are
the assumptions that the neutral ions trace the \hi, and
that emission measurements extending over the radio
telescope beam of ten
arcminutes can be safely compared to absorption measurements sampled over
an effectively infinitesimally small beam equal to the angular size of
the AGN. In light of these concerns, we estimate an error
of 5\kms\ in our \fuse\ velocities after alignment. A detailed
discussion of velocity calibration issues in \fuse\ spectra, and how 
to deal with them, is given in \citet{Wa03}.

Since each pixel on the \fuse\ detectors has a velocity width of
$\approx2$\kms, but the instrumental resolution is only $\approx20$
\kms\ (FWHM), the data from the instrument are highly oversampled, and 
so we have rebinned each spectrum by five pixels (i.e. to $\approx10$
\kms\ bins) to obtain the versions displayed and measured in this
paper. We fit continua to our spectra using low order ($n=1-3$) Legendre
polynomials over $\approx5$\,\AA\ regions on either side of each line
under study.

The primary focus of our investigation is \os\ $\lambda1031.926$,
since the presence of transition-temperature gas in high-velocity
clouds has only recently been discovered. Absorption
from \ion{C}{2}$^*~\lambda1037.018$ is responsible for blending with
high-velocity (Complex~C) absorption from the weak \os\ line, at
1037.627\,\AA; we therefore do not include measurements of the weak \os\
line in this work. We also study \ion{Si}{2} $\lambda1020.699$,
\ion{C}{2} $\lambda1036.337$, and \ion{Ar}{1} $\lambda1048.220$. The
\ion{C}{2} line is very strong, and although this typically leads to saturation
in the line center, the high sensitivity in the wings helps to reveal
the full extent of the neutral gas absorption; the \hi\ 21\,cm line is not
sensitive to these lower column density regions. Unfortunately,
we could make no clean observation of \ion{C}{3} $\lambda977.020$, a tracer
of intermediate-ionization gas, at Complex~C velocities, because the
\ion{C}{3} line is blended below $-100$\kms\ by zero velocity
absorption from \ion{O}{1} $\lambda976.448$. The same blending problem 
afflicts high-velocity absorption by \ion{Fe}{3} $\lambda1122.524$,
this time by \ion{Fe}{2} $\lambda1121.975$. Although \ion{C}{3}
or \ion{Fe}{3} are stronger and hence more preferable tracers of
intermediate-ionization gas, we do include profiles of
\ion{S}{3} $\lambda1012.495$, to trace the kinematics of this phase of
interstellar gas.

The strength of contamination from molecular hydrogen varied
between our five sight lines. The H$_2$ (6--0) P(3) line at
1031.191\,\AA\ is the offending line, since in the rest frame of \os\
$\lambda1031.926$ it has a velocity of $-213.5$\kms, near the negative
velocity limit of Complex~C. For Mrk~876, where the H$_2$ column
density is the highest, we corrected the \os\ profile for
contamination from H$_2$ by measuring the other H$_2$ lines in the
vicinity of the \os\ line. Our model of the H$_2$ (6--0) P(3) line
included two components (1: $v_{LSR}=-7$\kms, depth = 0.60, FWHM =
31\kms; 2: $v_{LSR}=-36$\kms, depth = 0.30, FWHM = 19\kms). The effect
of this process is to reduce the measured high-velocity \os\ column density
slightly. To account for the errors this process introduces, our
systematic error on the \os\ column density towards Mrk~876 includes
the effect of varying the model parameters: $v$(H$_2$) by $\pm10$\kms,
the H$_2$ depth by
$\pm10$\%, and the H$_2$ width by $\pm10$\%. Towards \pgb, the sight
line with the second highest molecular hydrogen column density, the
Complex~C \os\ absorption does not extend to $-200$\kms, so there
is no H$_2$ interference.

\subsection{HST}

The GHRS and STIS spectra we use are described in Table 3. We refer
the reader to \citet{Br94} and \citet{Ki98} for descriptions of the
on-orbit performance of the GHRS and STIS instruments, respectively. All data
were all taken with intermediate resolution gratings, and were reduced using
standard processing pipelines. The grating used and its resolution,
found in the GHRS \citep{So95} and STIS \citep{Pr02} Instrument
Handbooks, are listed for each exposure in columns three and eight.  

In order to achieve consistency in comparing \fuse\ and \hst\ spectra,
we chose to display all absorption line spectra in 10\kms\ bins.
Given that the pixel size between GHRS and STIS exposures varies
(column nine), our wish for 10\kms\ rebinned pixels implied a
different rebinning factor for the different \hst\ spectra, and this
factor is given in column ten of Table 3. 

The velocities were converted from the heliocentric to the LSR
reference frame using the corrections listed
in Table 1. The velocity calibration from the \hst\ detector pipelines is far
more reliable than from the \fuse\ pipelines, so no further
alignment was considered necessary. Any residual errors introduced by
inaccuracies in the GHRS and STIS wavelength scales are unlikely to
change the conclusions of this paper, since the kinematic
inter-relationships we focus on are mainly between absorption lines in the
\fuse\ dataset. 

We used $\approx10$\,\AA\ regions on either side of \nf\
$\lambda1238.821$ for continuum fitting, in the same manner as 
was used for the \fuse\ data (\S3.1). In the case of \pga, the
same procedure was used for  \sif\ $\lambda1393.755$, \cf\ $\lambda1548.195$, 
and various low and intermediate ionization lines. All our targets are
extragalactic sources, which tend (unlike stellar spectra) to have
flat continua, so the continuum placement process was generally
straightforward. 

\section{Results}

Our measurements of high-velocity, highly ionized absorption are
presented in Table 4. For each high ionization absorption line
detected at Complex~C velocities, we list the rest vacuum wavelength
($\lambda$), 
velocity range of high-velocity absorption ($v_{min}$, $v_{max}$),
velocity centroid and width of high-velocity absorption ($\bar{v}$ and
$b$), obtained through either the moments of the apparent optical
depth profile or through Gaussian component fitting, equivalent width 
($W_{\lambda}$), column density ($N$) and signal-to-noise ratio  
per resolution element (S/N) in the vicinity of the line. All 
velocities in this table, and throughout this paper, are quoted in the
LSR reference frame, and all column densities quoted are measured in the
high-velocity gas, not integrated over the full extent of the absorption. 
\ion{S}{3} is not usually recognised as a high ion, but we include
measurements of high-velocity \ion{S}{3} absorption here to assess the
relationship between absorption in intermediate and highly ionized gas. 

The velocity ranges we used for the Complex~C absorption were chosen
after careful consideration of the extent of the high-velocity
absorption in different species along each sight line. In some cases 
these are different (by up to 20\kms) from the
velocity ranges used in S03, due to the addition of newer, higher S/N
data and improvements in the pipeline processing.

Our table makes use of two methods for finding the central velocity
and width of 
highly ionized absorption. When possible, we fit Gaussian components
to the data, and present the formal central velocity ($\bar{v}$) and
width ($b$) of the
Gaussian. This is only possible when the high-velocity component is
distinct and free from blending with other lines. The other method
makes use of the moments of the apparent optical depth, found from the
continuum-normalized flux profile $F(v)$ by
\begin{equation}
 \tau_a(v)=\mathrm{ln}\frac{1}{F(v)}. 
\end{equation}
The first moment gives the velocity centroid 
\begin{equation}
\bar{v}=\int^{v_{max}}_{v_{min}}v\tau_a(v)\mathrm{d}v/\int^{v_{max}}_{v_{min}}\tau_a(v)\mathrm{d}v.\end{equation}
The second moment yields the Doppler width
\begin{equation}
b=\sqrt{2\int^{v_{max}}_{v_{min}}(v-\bar{v})^2\tau_a(v)\mathrm{d}v/\int^{v_{max}}_{v_{min}}\tau_a(v)\mathrm{d}v}.
\end{equation}
When the Gaussian fit method is available, we use this as the preferable
indicator of velocity centroid, since the moment method gives a skewed
result when the absorption is highly asymmetric about the line
center. With either method, our error on $\bar{v}$ accounts for
both statistical errors and uncertainty in the velocity zero point
(5\kms\ for \fuse\ spectra, and 2\kms\ for \hst\ data). The error on
$b$ accounts only for statistical errors, since the line
width is insensitive to the zero point of the velocity scale.

The column densities were calculated using the apparent optical depth (AOD)
technique \citep{SS91}; first $\tau_a(v)$ is converted to $N_a(v)$ using
\begin{equation}
N_a(v)=\left(\frac{m_ec}{\pi e^2}\right)\left(\frac{\tau_a(v)}{f\lambda}\right)=3.768\times10^{14}\left(\frac{\tau_a(v)}{f\lambda}\right),
\end{equation}
where $f$ is the oscillator strength and $\lambda$ the wavelength of
the transition (expressed in \AA\ when using the second form, to give
$N_a(v)$ in \sqcm\,(km\,s$^{-1}$)$^{-1}$). Equation
(4) is then integrated to find the total column density between two
velocity limits: 

\begin{equation}
N_a=\int_{v_{min}}^{v_{max}}N_a(v)\,\mathrm{d}v
\end{equation}
For sight lines in which we detect no high-velocity \ion{S}{3} or \nf\
absorption, we present 3$\sigma$ upper limits for the column density,
unless blending prevents such a measurement being made. 

There are two errors quoted for each measurement of equivalent width and column
density. The first is a statistical error, found from a quadrature
sum of uncertainties in the count rate (Poisson noise) and continuum
placement uncertainty. The second error is a conservative estimation of
systematic error, found from a quadrature addition of fixed pattern
noise in the detectors, and choice of velocity integration
limits. We estimated fixed pattern errors per pixel of
10\% for \fuse, 2\% for GHRS, and 1\% for STIS detectors, and an
uncertainty in the velocity limits of 15\kms\ for \fuse\ spectra, and 6
\kms\ for both GHRS and STIS spectra. Note that this velocity
error itself has two components: one due to the intrinsic uncertainty
in the post-calibration velocity scale, and one due to the uncertainty
over which velocities should be used to define high-velocity
absorption. It is the velocity range uncertainty which dominates our
systematic error in most cases, since the high-velocity absorption is
not always distinct from low-velocity absorption, and sometimes the
chosen division is somewhat arbitrary. 
The signal-to-noise ratio per resolution element is calculated by
measuring of the rms dispersion of the data around the fitted
continuum, in the proximity of each line.  
In every case our independently measured \os\ 
column density is statistically identical (within the 1$\sigma$
error) to the one previously published in S03. 

\section{Relationships between High Ion and Low Ion Absorption} 

Figures 2, 3, and 4 display stacks of absorption line and \hi\ 21\,cm
emission line profiles for Mrk~279 and Mrk~817, Mrk~876 and
\pgb, and \pga\  
respectively. Dashed vertical lines correspond to the peaks in the neutral
hydrogen emission line profile given in the top panel, and show the
velocities at which the neutral gas column density is highest. Dotted lines
illustrate our continuum placement, chosen using the technique described in
\S3. Absorption in the velocity range corresponding to Complex~C  
absorption is shaded in gray, according to the limits quoted in Table 4
-- we note that the width of the Complex~C absorption components may
vary between species, hence the shading is included principally for
illustrative purposes. Short vertical tick marks indicate the location of
absorption blends. For Galactic blends, the tick marks
designate the velocity where we expect the component to appear,
assuming the absorbing ion occurs at the same velocity as the neutral
hydrogen. In the cases of Mrk~876 and \pgb, many of the blends
have a two-component structure, which is marked accordingly. For extragalactic
blends, where the absorption blend occurs at higher redshift, the tick
marks simply designate the line center of the blend. 

\subsection{Mrk~279}
The \hi\ profile towards Mrk~279 shows eight components between $-150$
and 50\kms, including two Complex~C features at $-137$ and $-102$
\kms\ (left panel of Figure 2). There appears to be corresponding absorption
centered near $-137$\kms\ in the \ion{Si}{2}, \ion{C}{2}, \nf\ and
\os\ profiles. The detection of high-velocity \nf\ absorption,
although tentative, is particularly interesting, and was first
reported in \citet{Pe00}. This feature is centered at $-149\pm5$\kms,
and we measure its equivalent width as $17\pm5\pm3$\,m\AA, between the
velocity limits of $-190$ and $-115$\kms, corresponding to
log\,$N$(\nf)$=12.93^{+0.12~+0.02}_{-0.17~-0.01}$. Since the 
systematic errors change the equivalent width but not the presence of
the feature (i.e., it could be $14\pm5$ or $20\pm5$\,m\AA), the formal
significance of the high-velocity \nf\ detection is 
$17/5>3\sigma$. This absorption extends down to $-190$\kms, not to
$-210$\kms\ as do the other ions shown in Figure 2.
The large number of neutral hydrogen components present in this direction
makes the separation of low and high-velocity gas
difficult. However, it is striking how the high-velocity \os\
absorption traces the shape of the \ion{C}{2} absorption closely; on
the negative side of the high-velocity absorption, both
ions recover to the continuum just beyond $-210$\kms, and on the
positive side they both start to recover at $-115$\kms. Though the \ion{C}{2}
line is heavily saturated, damping wings would not appear unless
$N$(\ion{C}{2})~$\gtrsim10^{21}$\sqcm, so the observed wings are likely
kinematic, not Lorentzian. With our chosen velocity
limits, we measure $W_{\lambda}$(\os)$=53\pm6\pm7$\,m\AA, 
corresponding to log\,$N$(\os)$=13.66^{+0.04~+0.05}_{-0.04~-0.04}$.
Gaussian component fitting reveals the center of this absorption to be
$-133\pm6$\kms, which (within the error) is the same velocity as the
\hi\ line center ($-137$\kms).

\subsection{Mrk~817}
Complex~C gas is very pronounced in the \hi\ profile towards Mrk~817,
with a strong, distinct component centered at $-109$\kms\ (right panel
of Figure 2). This feature is mirrored in the \ion{Si}{2}, \ion{C}{2},
\ion{S}{3}, and \os\ profiles, which all display clear absorption at this
velocity. The \ion{S}{3} detection is the strongest among all the Complex~C
sight lines, with an equivalent width
$W_{\lambda}$(\ion{S}{3})$=32\pm3\pm10$\,m\AA, and a central velocity of
$-114\pm9$\kms, showing a kinematic connection between the neutral and
intermediate-ionization gas. We do not expect contamination from the
H$_2$ (0--0) P(2) $\lambda1012.173$ line, since there is no H$_2$ absorption
out of the $J=2$ levels in this direction.
A double intergalactic Ly$\alpha$ feature happens to occur at just the
redshift to blend with \nf\ $\lambda1238.821$, so although at first
glance \nf\ absorption appears to be present, we have no way of
knowing conclusively. Strong broadening from the IVC at
$-40$\kms\ blends with high-velocity \os\ absorption, so the true extent of
high-velocity \os\ is hard to ascertain, but with a velocity range
choice of $-160$
to $-80$\kms\ we obtain $W_{\lambda}$(\os)$=93\pm2\pm29$\,m\AA,
and hence log\,$N$(\os)$=13.97^{+0.02~+0.08}_{-0.02~-0.11}$. The mean
\os\ velocity is $-109\pm10$\kms, exactly the same as the \hi\
centroid. What appears to be a positive velocity wing in 
the \os\ profile is actually an intergalactic Ly$\beta$ absorber at
$2360\pm330$\kms, a velocity which suggests an association with the
Canes Venatici galaxy grouping \citep{Tu88}.

\subsection{Mrk~876}
The Mrk~876 sight line (left panel of Figure 3) has a two-component
high-velocity 
\hi\ profile, with a very weak $-173$\kms\ component and a stronger
$-133$\kms\ component, seen weakly in \ion{Si}{2} and more clearly
in \os. The \ion{Ar}{1} and \nf\ lines show no significant
high-velocity detection. \ion{C}{2} $\lambda1036.337$ is too saturated
to make measurements near the line center, but reveals the presence of
absorption out to $-220$\kms. High-velocity \ion{S}{3} absorption is hard
to separate from the strong molecular hydrogen absorption in
this direction. The high-velocity \os\ absorption is extended and
broad, with no recovery to the continuum out to $<-220$\kms, at which
velocity a blend with H$_2$ (6--0) P(3) $\lambda1031.191$
becomes significant. We
measure the velocity centroid of \os\ to be $-148\pm9$\kms, which is
only different from that of the \hi\ component at the 2$\sigma$ level;
given the strong blending from H$_2$ lines, we cannot say whether this
difference is real. With
log\,$N$(\os)$=14.12^{+0.02~+0.09}_{-0.02~-0.11}$, Mrk~876 has the
highest high-velocity column density of \os\ of any Complex~C sight
line in our sample, yet with log\,$N$(\hi)$=19.39\pm0.02$
the lowest neutral hydrogen column density.

\subsection{\pgb}
The \pgb\ sight line has strong high-velocity \hi\ emission at
$-158$\kms, and a weaker component at $-115$\kms\ (right panel of
Figure 3). Absorption at
this velocity is not clear in \ion{Ar}{1}, but clearly exists in
\ion{Si}{2} and \ion{C}{2}. No high-velocity gas is apparent in
\ion{S}{3}; the high column density of molecular hydrogen along
this sight line causes strong blending between H$_2$ P(4) (8--0)
$\lambda1012.261$ and the \ion{S}{3} line and makes an assesment of the
presence of high-velocity \ion{S}{3} absorption difficult. Note that
the molecular hydrogen lines in our plots have a two-component nature,
due to local and IVC absorption. Absorption is
present in \os, but its extent is hard to ascertain, because of broad local 
absorption near 0\kms, and blending with H$_2$ (6--0) P(3) 
$\lambda1031.191$ below $-200$\kms. Integrating over the optical
depth profile reveals an \os\ central velocity of $-147\pm10$\kms, 11\kms\
redward of the \hi.
We measure an upper limit of log\,$N$(\nf)$<13.39$ and
log\,$N$(\os)$=13.66^{+0.08~+0.11}_{-0.08~-0.15}$, among the lowest
columns of high-velocity \os\ in our five 
sight lines, 0.46 dex less than the value towards Mrk~876. We also note the
presence of an extended positive velocity wing in \os, the like of which has
been noted by S03 in 22 sight lines, 18 of which lie in the
Northern Galactic hemisphere. Although apparently unrelated to Complex
C gas, these wings nonetheless constitute a tracer of highly ionized
high-velocity gas whose origin is unknown.

\subsection{\pga}
Two stacks of absorption lines for \pga\ are presented in
Figure 4, since in this case we have the benefit of STIS E140M
data with extensive wavelength coverage. We include the \hi\ and \os\
profiles in both panels for ease of comparison.
The structure of the nearby absorption along the \pga\ sightline 
is clearly illustrated in the \hi\ profile, which shows
three distinct components: zero-velocity absorption, presumably tracing
nearby gas, intermediate-velocity gas (centered at $v=-54$\kms) and
high-velocity gas (centered at $v=-128$\kms). These
components are clearly mirrored in absorption in \ion{Si}{2}
$\lambda1020.699$ and \ion{Ar}{1} $\lambda1048.220$, and trace the
densest regions of gas along the sight line. An extra
absorption component along this sight line at $v=-110$\kms\ was
suggested by \citet{Ri01a}, upon detailed analysis of STIS data, and has been
confirmed by component fitting of the \ion{O}{1} $\lambda1302.169$ line
\citep{Se03b}. \ion{S}{3} is detected in Complex~C with an equivalent width
$W_{\lambda}$(\ion{S}{3})$=17\pm5\pm12$\,m\AA. 

With regard to the highly ionized Complex~C gas, components of
absorption are clearly seen at Complex~C velocities in \sif, \cf\ and
\os. In each of these three cases, the weaker member of the resonance
doublet was of no value in measuring a column density, due to either a lack
of discernable absorption, or contamination from other lines. We measure
log\,$N$(\sif)$=12.73^{+0.02~+0.03}_{-0.01~-0.02}$,
log\,$N$(\cf)$=13.26^{+0.03~+0.01}_{-0.04~-0.02}$, and 
log\,$N$(\os)$=13.71^{+0.04~+0.05}_{-0.04~-0.05}$. High-velocity
\nf\ absorption is not seen in the data; we measure a 
$3\sigma$ upper limit to the high-velocity \nf\ column density of log
$N$(\nf)$<12.85$. This non-detection is qualitatively consistent
with the low N/O ratio measured by \citet{Ri01a} and \citet{Co03a} in
this direction. The kinematics of the high-velocity absorption are
interesting, since we measure $\bar{v}$(\sif)$=-119\pm3$\kms,
$\bar{v}$(\cf)$=-106\pm4$\kms, and $\bar{v}$(\os)$=-110\pm5$\kms, all by
component fitting. This sight line thus exhibits a $18\pm6$\kms\ 
difference between the centers of velocity of \hi\ and \os, with the
\cf\ appearing to follow the \os, and the \sif\ falling inbetween. The
intermediate-ionization \ion{S}{3} line has a velocity centroid of
$-104\pm9$\kms. Note that absorption from \ion{Si}{2}, \ion{Si}{3}, 
and \sif\ is seen over the same velocity range.
 
To further illustrate the kinematic structure of the Complex~C gas
toward \pga, we compare apparent column density profiles of various
species in Figure 5. In each panel one of the profiles has been
normalized in order to compare its shape with the other profile. We
include ions that trace low-ionization gas (\hi\ and
\ion{Si}{2}), intermediate ionization gas
(\ion{S}{3}), and highly ionized gas (\sif, \cf, and
\os). Complex~C absorption is clearly seen in all these ionic
species. The two component high-velocity structure is most easily seen
in the second panel, where we compare \ion{Si}{2} $\lambda1526.707$
with \cf\ $\lambda1548.195$ from the STIS data. We return to a
discussion of the absorption kinematics in \S8.2.

\section{Potential Ionization Mechanisms}

In studying highly ionized gas in HVCs our ultimate goal is to
determine the ionization mechanism(s). Not only is this interesting in
its own right, but also knowledge about ionization may be able to
provide information on the location of HVCs. 
In this section we describe the potential mechanisms that could create
highly ionized gas in or around HVCs.

\subsection{Collisional Ionization Equilibrium (CIE)}

Perhaps the simplest picture that can account for simultaneous
observations of \sif, \cf, and \os\ (as seen toward \pga) is
one involving a cloud of gas at a fixed kinetic temperature, with every
collisional ionization balanced by a radiative
recombination. \citet{SD93} calculated the fraction of each element in
each ionization stage, assuming these equilibrium conditions. To test to see
if CIE conditions exist, one needs to find that the column densities
of multiple ions are simultaneously consistent with the model
predictions at one temperature. However, in interstellar and intergalactic
environments, such conditions rarely apply to the \os\ ion, since it
exists at temperatures of a few\,$\times10^5$\,K where the cooling function is
maximized. When cooling occurs faster than recombination the
ionization can become ``frozen-in'', producing gas referred to as
overionized \citep{Ka73}. A single phase of hot gas under
CIE conditions would be traced by absorption at the same central velocity in all
ions. Velocity offsets cannot be explained by a single-phase,
single-temperature CIE model.

\subsection{Photoionization} 

The high ions \sif, \cf, \nf, and \os\ require energies of
33.5, 47.9, 77.5, and 113.9\,eV respectively for their creation. If
the radiation field is hard enough to contain significant intensities
of photons at these energies, the ions can be produced directly by
photoionization. S03 have conducted photoionization models to investigate
the ionization fraction of \os\ ($f_{O~VI}$) in HVCs as a function of
neutral hydrogen column density and ionization parameter
$U=n_{\gamma}/n_H$, where $n_{\gamma}$ and $n_H$ are the photon
density and total hydrogen density, respectively. The models assumed the gas is
optically thin to ionizing and 
cooling radiation (i.e. log $N$(\hi)$<17.2$), and took the QSO 
spectral energy distribution (SED) from \citet{Ma92}, normalized at 912
\,\AA\ to
$J_{{\nu}_0}=1\times10^{-23}$\,erg\sqcm\,s$^{-1}$\,Hz$^{-1}$\,sr$^{-1}$
\citep{HM96}. The S03 photoionization models provides a strong
conclusion: to produce an \os\ column density of 10$^{14}$\sqcm\
requires the HVC to be several hundred kiloparsecs in depth, 
for any metallicity in the range $Z=0.01$--$1.0Z_{\odot}$. These sizes
are far too large to be consistent with most
models of HVCs; \citet{Bl99} predict typical diameters of $\sim25$\,kpc;
theories that place HVCs closer to the Milky Way predict even smaller sizes.
If the \hi\ column density in the model is increased
above log\,$N$(\hi)$<17.2$, so that the cloud is no longer optically
thin, the predicted \os\ and \cf\ column densities drop, and the conclusion
becomes even stronger: \os\ is most likely collisionally ionized. 

More detailed photoionization modeling of the high ions in Complex~C
would need to include the effect of ionizing photons
escaping from the Galactic disk. A key uncertainty in doing this is
the escape fraction of ionizing photons \citep{BM99}.  The strong
\ion{He}{2} ionization edge at 54 eV in hot 
star spectra limits the flux of photons energetic enough to produce
either \nf\ or \os. Therefore, in studies of highly ionized gas in the
ISM of the Milky Way, \nf\ and \os\ are almost always interpreted as
being tracers of collisionally ionized gas \citep[e.g.][]{Sa03, Fo03}. 
So, modifications to the radiation field will be more significant for
calculating the ion fractions of low and intermediate ionization species,
than for high ions such as \os.

\subsection{Radiatively Cooling Gas Flows (RC)}

As hot ($T_0\gtrsim10^6$\,K) gas cools down by radiative recombination
it will pass through the temperature regime at which high ions such as
\nf\ and \os\ exist.
\citet{EC86} estimated the ionization properties of a body of gas cooling
from an initial high temperature ($T=10^6$\,K), for a variety of
cases ranging from isochoric (constant density) to isobaric (constant
pressure). Their models predict the column density of \sif, \cf, \nf,
and \os; for the case of \os, the column density is given by 
$N$(\os)~$=4-6\times10^{14}(v_{cool}/10^2$\kms)\sqcm, assuming solar
abundances, where $v_{cool}$ is the flow velocity.
\citet{He02}, upon discovering a correlation between line width and
column density in \os\ absorbers across a wide variety of locations,
have suggested that all \os\ absorbers can be explained by radiatively 
cooling hot gas passing through transition temperatures. This
conclusion remains the same over a wide range of metallicity. This
result relies on there being a linear relationship between the
characteristic flow velocity and the non-thermal line broadening
parameter $b_{nt}$. 

If the RC model is to give a complete description of the highly
ionized gas present in or near HVCs, it would need to
account for the origin of the hot gas in the first place, and the
kinematical alignments between neutral and ionized phases of gas that
are reported in this paper.

\subsection{Conductive Interfaces (CI)} 

When reservoirs of hot and cold gas come into contact with one
another, electron collisions will conduct energy from the hot medium
towards the cooler medium, at a rate depending on the plasma
conductivity and the orientation of the magnetic field relative to the
interface. Transition-temperature ions, such as \cf, \nf, and \os, can
be produced in the conduction front as the cooler gas evaporates and
the hotter gas condenses. 

\citet*{Bo90} studied the energy transport when
a planar front is established between an interstellar cloud and a hot
($T\sim10^6$\,K) coronal medium. The calculations of \citet{Bo90}
predict the expected column
densities of the ions \sif, \cf, \nf, and \os\ produced in a
conductive interface, as a function of both time and the angle $\theta$ between 
the normal to the front and the magnetic field. We consider magnetic
field orientations of $\theta=0\degr-85\degr$ and interface ages of
log\,[$t$(yr)]~$=5.0-7.0$. Since the thermal conductivity of plasma is
far greater
along magnetic field lines than across them, the conduction is
quenched when the field runs parallel to the interface, and so the
predicted high ion column densities are much lower in this case. As
with any interface model, the CI theory predicts that there should be no 
(or small) velocity offsets between high ions observed at the surface of the
cloud and neutral ions observed within it. Furthermore, the line
widths are predicted to be thermally dominated. These conditions
provide a set of diagnostic tests that can be applied when trying to
explain the origin of an \os\ absorber.

\subsection{Turbulent Mixing Layers (TML)}

Turbulence is increasingly being considered to be a ubiquitous phenomenon
in the ISM. In an interface between turbulent hot gas and a region of
cold gas, any relative motion of the gas phases can generate
Kelvin-Helmholtz instabilities that mix the hot and cold gases together
in a mixing layer. Such a layer would be at transition temperatures,
sufficiently hot to generate ions such as \nf\ and \os\ by
electron collisions. TML theory was introduced by \citet{BF90}, who
argued that conduction will only form a significant mechanism of
energy transport in a quiescent medium, but in the turbulent hot ISM
tangled magnetic fields will suppress conduction. \citet*{Sl93}
developed the TML theory to predict high ion column densities, as a
function of the entrainment velocity and mixing-layer
temperature.

We consider the TML predictions with the entrainment velocity of the
gas flow in the range $25-100$\kms, and the temperature of the
post-mixed gas in the range log\,$T=5.0-5.5$. A key difference between
TML theory and other ionization models is a high $N$(\cf)/$N$(\os)
ratio; observation of this quantity thus provides a key test of the
TML theory. 

\subsection{Shock Ionization (SI)} 

If a high-velocity cloud passes through a surrounding medium at a
velocity higher than the local sound speed, a shock front will develop
at the leading edge of the cloud. The temperature in the gas behind
the front could then be raised to levels at which highly ionized
material can be formed by collisional ionization ($T\gtrsim10^5$\,K).  
\citet{DS96} generated a grid of low-density shock
models, predicting the column density of high ions for shock
velocities in the range $150<v_S<500$\kms, assuming solar
abundances. In this study we consider the SI model with
shock velocities of $150-500$\kms\ and magnetic parameters
$0\le B_0/n_0^{3/2}\le 4\,\mu\mathrm{G\,cm}^{-3/2}$.  

In the shock ionization model, the shock velocity is
the key parameter determining the expected column densities of high
ions. We note that since Complex~C is falling toward the Galaxy (at
$\approx130$\kms), any Galactic wind or fountain outflow would
serve to increase the Mach number and hence the strength of the shock. In
this scenario, the interaction of the cloud with the surrounding
environment would be contributing to both the ionization and the
metallicity of the cloud gas. Instabilities arising from the
interaction of a HVC with magnetic fields could also influence the
ionization properties of the cloud \citep{Ko01}.

\subsection{Supernova Remnants (SNR)}

\citet{SC92} and \citet{Sh98} have considered the evolution of low
density supernova remnants expanding into a surrounding medium. However,
since no stars have ever been detected in HVCs despite extensive
searches \citep{Hp03, Wi02, Da02}, we do not expect them to
harbor SNRs, and so we do not consider the SNR model any further. 

\section{Non-Solar Abundance Issues}

A difficulty we face in applying ionization models to Complex~C is the
fact that the elemental abundance ratios are known to be
non-solar in the complex. In this section we investigate the effect that
changed abundances will have on the model high ion column density
predictions. 

In any model where the predicted columns of high ions are regulated by
radiative cooling losses, the \os\ column densities are fairly
insensitive to metallicity. This is because oxygen is the primary
coolant for interstellar gas with log $T=5.0-5.6$, and 
so reducing the oxygen abundance reduces the cooling rate, and hence
lengthens the cooling time of the gas. This feedback effect causes 
the predicted columns to be fairly constant over metallicity
\citep{EC86, He02}. \citet{Sa03} discuss how the expected value of $N$(\os) 
depends on the metallicity in the cooling gas of a Galactic fountain
flow. Since the total column density of cooling gas is given by
$N_{cool}=\dot{N}_{\mathrm{H}}t_{cool}$, where $\dot{N}_{\mathrm{H}}$
is the flux of cooling gas (hydrogen ions\sqcm\,s$^{-1}$) and
$t_{cool}$ is the cooling time of the gas, the expected column density
of \os\ is given by 

\begin{equation}
N(\mathrm{O~VI})=f_{\mathrm{O~VI}}A_{\mathrm{O}}\dot{N}_{\mathrm{H}}t_{cool},
\end{equation}

\noindent where $f_{\mathrm{O~VI}}$ is the fraction of oxygen atoms
present as \os, and $A_{\mathrm{O}}$ is the abundance of oxygen with
respect to hydrogen. 

It is often assumed for departures from solar abundances
that $t_{cool}\,{\propto}\,n_{\mathrm{H}}^{-1}A_{Z}^{-1}$, where
$n_{\mathrm{H}}$ is the initial ionized hydrogen density in the 
cooling gas and $A_Z$ is the mean metallicity of the dominant coolants
in the gas. This relation assumes that the cooling is dominated by the
radiative emission produced by the heavy elements. Since oxygen is the
most important coolant, we take $A_Z=A_{\mathrm{O}}$, and therefore
$N$(\os)$\,\propto
f_{\mathrm{O~VI}}\dot{N}_{\mathrm{H}}n_{\mathrm{H}}^{-1}$, which is
independent of 
the oxygen abundance. Since $\dot{N}_{\mathrm{H}}n_{\mathrm{H}}^{-1}$ has
dimensions of velocity, we can identify
$\dot{N}n_{\mathrm{H}}^{-1}\equiv v_{cool}$, so that
$N$(\os)$\,\propto f_{\mathrm{O~VI}}v_{cool}$.

However, the calculations of \citet{Be01} show that the cooling time
scales more like
$t_{cool}\,{\propto}\,n_{\mathrm{H}}^{-1}A_{Z}^{-\beta}$, where
$\beta\approx0.85$ over the range $Z=0.1-1.0Z_{\odot}$. Physically,
this effect is due to the inclusion of cooling from hydrogen and
helium as well as metal line cooling. In this case,
the expected column density of \os\ will scale as
 
\begin{equation}
N(\mathrm{O~VI})\,\propto A_{\mathrm{O}}^{1-\beta}v_{cool}
%\approxA_{\mathrm{O}}^{0.15}v_{cool}.
\end{equation}

\noindent So, if the abundance of oxygen changes from
$A_{\mathrm{O}}^{\odot}$ to 
$A_{\mathrm{O}}^{\prime}$, then the predicted column density of
\os\ will change from $N$(\os)$^{\odot}$ to $N$(\os)$^{\prime}$ according to

\begin{equation}
\frac{N(\mathrm{O~VI})^{\prime}}{N(\mathrm{O~VI})^{\odot}}
=\left(\frac{A_\mathrm{O}^{\prime}}{A_{\mathrm{O}}^{\odot}}\right)^{1-\beta}.
\end{equation}

For elements other than oxygen, which are less important coolants, the
predicted column densities will also scale as
$N(\mathrm{X})^{\prime}/N(\mathrm{X})^{\odot}=(A_{\mathrm{O}}^{\prime}/A_{\mathrm{O}}^{\odot})^{1-\beta}$, 
{\it provided that}
$A_\mathrm{X}^{\prime}/A_{\mathrm{X}}^{\odot}=A_\mathrm{O}^{\prime}/A_{\mathrm{O}}^{\odot}$.
However, if there are deviations from the solar ratios among the other
heavy elements, then 

\begin{equation}
\frac{N(\mathrm{X})^{\prime}~}{N(\mathrm{X})^{\odot}}=
\frac{(A_{\mathrm{X}}/A_{\mathrm{O}})^{\prime}~(A_\mathrm{O}^{\prime})^{1-\beta}}
{(A_{\mathrm{X}}/A_{\mathrm{O}})^{\odot}(A_{\mathrm{O}}^{\odot})^{1-\beta}}=
\frac{A_{\mathrm{X}}^{\prime}}{A_{\mathrm{X}}^{\odot}}
\left(\frac{A_\mathrm{O}^{\prime}}{A_{\mathrm{O}}^{\odot}}\right)^{-\beta},
\end{equation}

\noindent where X could represent either \sif, \cf, or \nf. Combining equations (8)
and (9) leads to the shift applied to the logarithmic column density ratios

\begin{equation}
\mathrm{log}\left[\frac{N(\mathrm{X})}{N(\mathrm{O~VI})}\right]^{\prime}\!=
\mathrm{log}\left[\frac{N(\mathrm{X})}{N(\mathrm{O~VI})}\right]^{\odot}\!+
\Delta(\mathrm{log}~A_{\mathrm{X}})-\Delta(\mathrm{log}A_{\mathrm{O}}),
\end{equation}

\noindent where
$\Delta(\mathrm{log}\,A_\mathrm{X})=\mathrm{log}\,A_\mathrm{X}^{\prime}-\mathrm{log}\,A_{\mathrm{X}}^{\odot}=\mathrm{log}(A_\mathrm{X}^{\prime}/A_{\mathrm{X}}^{\odot})$
is the change in the logarithmic abundance of element X. It can be
seen from Equation 10 that the adjusted logarithmic column density ratios are
independent of the exact value of $\beta$. 
 
These arguments were originally formulated
for the case of pure radiatively cooling gas flows (see \S6.3), but
equally apply to all other models considered in \S6.3--\S6.6,
since the rate of cooling is a critical 
component of all the other models. It should be emphasized that these
results are approximate and neglect the contribution to the cooling by
other elements (particularly carbon and nitrogen). Nonetheless, this
is a first attempt at quantifying the effects of low metallicities on
high ion production. Equation 10 has to be applied twice in 
order to apply the model ionic ratio predictions to the Complex~C gas;
these two corrections are described in the next two subsections.

The one model where cooling is not important is the CIE model (\S6.1), since
in this scenario the temperature is held fixed, and the equilibrium
ionization balance calculated. In this case, the expected column
density of a given ion is proportional to the fraction of the element
in the relevant ionization stage \citep[tabulated in][]{SD93}, multiplied by
the abundance of the element. 

\subsection{Updates to Solar Abundance Ratios}

The models described in \S6 each assumed different solar
elemental abundance ratios, according to the most up-to-date abundances
that were available when the models were published. However, the solar
abundances have recently undergone several important revisions, as can
be seen in Table 5, in which we summarize all significant published
values for the solar composition. We
made a correction so that each ionization model is adjusted to the same set of 
solar abundances. We decided to take the solar oxygen abundance
(log\,$A_{\mathrm{O}}^{\odot}=-3.31$) from \citet*{AP01}, the solar carbon
abundance (log\,$A_{\mathrm{C}}^{\odot}=-3.61$) from \citet*{AP02},
and the solar
nitrogen and silicon abundances (log\,$A_{\mathrm{N}}^{\odot}=-4.07$;
log\,$A_{\mathrm{Si}}^{\odot}=-4.46$) from \citet{Ho01}. These abundances are
significantly different (up to 0.2 dex) from the widely used tables of
\citet{AG89}. By finding the difference $\Delta A_{\mathrm{X}}$
between the solar abundances used in each model calculation and our adopted
solar abundances, we brought all model predictions onto a
unified solar abundance scheme, using equation (10) in the form  

\begin{equation}
\mathrm{log}\left[\frac{N(\mathrm{X})}{N(\mathrm{O~VI})}\right]^{\odot}_{new}\!\!\!\!=
\mathrm{log}\left[\frac{N(\mathrm{X})}{N(\mathrm{O~VI})}\right]^{\odot}_{old}\!\!\!\!+
\Delta(\mathrm{log}\,A_{\mathrm{X}}^{\odot})-
\Delta(\mathrm{log}\,A_{\mathrm{O}}^{\odot}).
\end{equation}

\noindent where the shifts
$\Delta(\mathrm{log}\,A_{\mathrm{X}}^{\odot})$ are in the sense ``new
minus old''.  

\subsection{Corrections for Environments with Non-Solar Abundance Ratios}

The second correction we make is to account for the fact that relative
abundance ratios in Complex~C are known to be non-solar
\citep{Wa99, Ri01a, Gi01, Co03a, Tr03, Se03b}. Caution is necessary when
interpreting various measurements of [X/H], since this number
depends upon the assumed solar elemental abundances. In Table 6 we
have taken all recent measurements of the metallicity of Complex~C and
adjusted them to our adopted abundances presented in Table 5. It can
be seen that all measurements of [O/H] in Complex C converge to a 0.41 dex
region between $-0.86$ to $-0.45$; toward \pga\ the updated
measurements all lie between $-0.86$ and $-0.79$. Given the errors 
there is no strong evidence for any variation in these abundances. The
\ion{O}{1}/\hi\ ratio is a sensitive measure of the overall
metallicity of the system, for a 
number of reasons. The depletion of oxygen onto dust is small 
\citep[e.g.][]{Mo02}, ionization effects are minimal since the
ionization potentials of \ion{O}{1} and \hi\ are very similar, and 
charge-exchange reactions couple the elements together \citep{FS71}. 
The latter two effects ensure that the ionization fractions of O and H 
track each other very closely. We now wish to determine the best
estimates of [X/H] in Complex~C for each of the elements
oxygen, nitrogen, silicon, and carbon so as to adjust the model predictions
accordingly. For this purpose we use the \citet{Se03b} measurement of
$N$(\hi)~$=19.94\pm0.06$ in the main component of Complex~C toward \pga. 

For oxygen, we use the most recent determination along the \pga\ sight
line of [O/H]$_C=-0.79$ from \citet{Se03b}, who used the same set of 
solar abundances as our adopted values.

For nitrogen we use log\,$N$(\ion{N}{1})~$=14.02$ in the Complex~C gas
toward \pga\ \citep{Co03a}. We also measure the high-velocity
\ion{N}{2} $\lambda1083.994$ towards \pga\ to have an equivalent width
of $111\pm4\pm25$\,m\AA\ between $-160$ and $-80$\kms, which using the AOD
technique corresponds to log\,$N$(\ion{N}{2})~$=14.24\pm0.02\pm0.14$.
We have no way of measuring $N$(\ion{H}{2}) in 
Complex C, which would allow us to explicitly calculate
[N/H]~=~[\ion{N}{1}+\ion{N}{2}]/[\hi+\ion{H}{2}]. However, we use the
results of \citet{Co03a}, who used photoionization modelling
to find that the ionization correction for
nitrogen is negligible, so that [\ion{N}{1}/\hi]~$\approx$~[N/H]. This
result implies that essentially the \ion{N}{1} and \hi\ reside
together in the neutral gas, and the \ion{N}{2} and \ion{H}{2} reside
together in the ionized gas, as might be expected since the first ionization
potentials of hydrogen and nitrogen are so close. Combining 
the $N$(\ion{N}{1}) measurement with $N$(\ion{H}{1}) leads to [N/H]$_C=-1.85$. 

For silicon, we utilize the \citep{Co03a} measurement of
log\,$N$(\ion{Si}{2})~$=14.67$ toward \pga, so that
[\ion{Si}{2}/\hi]~$=-0.81$. These authors found that the Si ionization correction
[\ion{Si}{2}/\hi]~$-$~[Si/H]~$\approx0.2$\,dex for
log\,$N$(\hi)~$=19.94$. With this assumption we derive [Si/H]$_C=-1.01$.

Unfortunately, for the Complex~C gas, no 
unsaturated absorption lines of carbon are available in either
the \fuse\ or \hst\ bandpass, from which we could make a carbon
abundance measurement, and the intersystem \ion{C}{2} transition at
2325\AA\ is too weak to measure. So, we decided to estimate the
relative carbon abundance in Complex~C by examining the behavior of
C/O versus metallicity as reported in the literature. In general, C/O
is known to be an increasing 
function of O/H at high metallicity, but is much flatter at low
metallicity \citep*{He00, Ge95}. The C/O ratio appears to plateau at
log\,(C/O)~$\approx-0.65\pm0.1$ in the \citet{He00} data for low
metallicity (log\,$A_{\mathrm{O}}<-4$) extragalactic \ion{H}{2}
regions and stars, and although Complex~C could have a different
nucleosynthetic history, we assume this ratio applies there. Combining this
estimate with the other abundance measurements discussed above we adopt  
\begin{eqnarray} 
\mathrm{(C/O)}_C =& 0.45\mathrm{(C/O)}_{\odot}\;\; &
\Rightarrow \mathrm{[C/O]}_C=-0.35\nonumber\\
\mathrm{(N/O)}_C =& 0.09\mathrm{(N/O)}_{\odot}\;\; &
\Rightarrow \mathrm{[N/O]}_C=-1.06     \\
\mathrm{(Si/O)}_C=& 0.60\mathrm{(Si/O)}_{\odot}\;\; &
\Rightarrow \mathrm{[Si/O]}_C=-0.22\nonumber.
\end{eqnarray}

\noindent Finally, since $\Delta(\mathrm{log}\,{A_{\mathrm{X}}})=$~[X/H] and
$\Delta(\mathrm{log}\,A_{\mathrm{O}})=$~[O/H], 
we can combine these abundances with equation (10) to generate the
relation used to correct the solar model column density predictions to apply
in Complex C.
\begin{eqnarray}
\mathrm{log}\,\left[\frac{N(\mathrm{X})}{N(\mathrm{O~VI})}\right]_{C}&=&
\mathrm{log}\,\left[\frac{N(\mathrm{X})}{N(\mathrm{O~VI})}\right]^{\odot}_{new}\!\!+
[\mathrm{X}/\mathrm{H}]_C-[\mathrm{O}/\mathrm{H}]_C\nonumber\\
&=&\mathrm{log}\,\left[\frac{N(\mathrm{X})}{N(\mathrm{O~VI})}\right]^{\odot}_{new}+
[\mathrm{X}/\mathrm{O}]_C
\end{eqnarray}

\noindent 
%Note that since [O/H]$_C=-1.00$, the predicted \os\ columns are reduced 
%from the solar abundance case by a factor $0.1^{0.15}=0.71$. 
Equations (11) and (13) were used to generate the modified
ratio predictions contained in Table 7, for each of the models
described in \S6.3--6.6.

All model predictions discussed here have taken no
account of elemental depletion into dust, which would further
complicate the interpretation of the ionic ratios. However, \citet{Co03a}
report solar [S/O] and [Si/O] ratios in Complex~C, suggesting that
dust depletion is not significant. This conclusion is reinforced by
strong detections of \ion{Fe}{2} in Complex~C \citep{Mu00, Tr03},
who found that the measured iron abundances leave little room for
depletion into dust. The absence of H$_2$ in Complex~C
\citep{Ri01b} is consistent with the absence of dust since H$_2$ is
thought to form on the surface of dust grains.

Finally, note that (with the exception of carbon) we have used
measured abundance ratios of the {\it neutral} gas in Complex~C; if
this HVC is interacting with a surrounding medium, then 
this medium could also display non-solar abundance patterns. In
any model where gas is mixed between two media, non-solar abundance
variations in the hot medium become a further complication, which is
not accounted for here.

\section{Comparison of Observations and Theory} 

\subsection{Ionic Ratios}

We focus our discussion of ionization on the high ions \sif, \cf, \nf,
and \os. The lower ion stages are more likely to be produced by
photoionization, since there are many more ionizing photons below
50\,eV capable of producing them. This origin has been suggested as an
explanation for low ionization species in other
high-velocity absorbers \citep[e.g.][toward PKS~2155-304 and Mrk~509]{Co03b}.

The best data available exist for the \pga\ sight line,
for which we measure high-velocity ionic ratios of
$N$(\sif)/$N$(\os)~$=0.10\pm0.02$,
$N$(\cf)/$N$(\os)~$=0.35^{+0.05}_{-0.06}$, and
$N$(\nf)/$N$(\os)~$<0.07$ (3$\sigma$). Our other new results are
measurements of the $N$(\nf)/$N$(\os) ratio: $0.19^{+0.06}_{-0.07}$
for Mrk~279, 
$<0.11$ for Mrk~876, and $<0.35$  for \pgb. Our ionic ratio
information is  displayed graphically in Figures 6 and 7, which 
show log\,[$N$(\cf)/$N$(\os)] versus log\,[$N$(\nf)/$N$(\os)],
and log\,[$N$(\sif)/$N$(\os)] against log\,[$N$(\nf)/$N$(\os)]
respectively. We choose to compare 
high ion column density ratios, rather than absolute column densities,
to allow for multiple interfaces or regions along the line of sight
where hot gas could exist. Using the model predictions summarized in
Table 7, we can 
identify regions of this ionic ratio space that gas behaving according
to the various models should occupy. Because of the various free
parameters that can vary in each model, we assume the models to occupy
boxes rather than lines in these ionic ratio diagrams. 
For each of four theories (CI, RC, TML, SI) we have computed the model
predictions for the cases of solar abundances (light regions) and
Complex~C abundances (darker regions). The changes in the predicted
ion ratios due to these abundance variations are fairly
substantial. For ease of comparison, blue arrows connect the two boxes for each
model. 

It can be seen that the adjusted RC model significantly underestimates the
observed $N$(\cf)/$N$(\os), $N$(\sif)/$N$(\os), and $N$(\nf)/$N$(\os)
ratios. The CIE model is incapable 
of simultaneously matching either the $N$(\cf)/$N$(\os) or the
$N$(\sif)/$N$(\os) ratio, in either abundance case. The SI model can
successfully reproduce both the  $N$(\cf)/$N$(\os) or the
$N$(\sif)/$N$(\os) ratios, but significantly underestimates the
$N$(\nf)/$N$(\os) ratio toward Mrk~279 by over an order of
magnitude. This narrows down the viable models to the CI and TML
theories. The CI model can reproduce the $N$(\cf)/$N$(\os) ratio
toward \pga, but seems to underestimate the $N$(\sif)/$N$(\os) ratio
along this sight line. The TML theory has the opposite problem: it can
match the $N$(\sif)/$N$(\os) ratio but overestimates
$N$(\cf)/$N$(\os). Both these theories 
(when adjusted for Complex~C abundances) predict $N$(\nf)/$N$(\os) ratios that
are marginally lower than the value observed toward Mrk~279, but they do
agree within the 2$\sigma$ error on the measurement. 
Note that although we have fully accounted for
errors in the data, we have not fully accounted for errors in the
model corrections. Uncertainties of $\sim0.2$\,dex are likely
because of incomplete knowledge of the elemental abundances; perhaps a
similar error arises from the assumption that the cooling is dominated
by oxygen. From these two plots we conclude that TMLs
or CIs could be potential explanations for the highly ionized material
around Complex~C. 
We have outlined these two model boxes in blue in Figures 6 and 7 to
highlight their position. 

Examining the CI model further, we note that the large size of the
CI model box in the ionic ratio planes is largely
due to the inclusion of interface age as a free
parameter. As we have no way of knowing the age of an interface around
Complex~C, we chose to include a range from $10^5$ to $10^7$ years, a
period which spans the transition from evaporative to condensing
interfaces. The data are most consistent with {\it young} ($\sim10^5$\,yr)
interfaces, since the adjusted-abundance CI boxes in Figures 6 and 7
overlap the data at their high $N$(X)/$N$(\os) extents, which
correspond to lower interface age. Physically this is because the \os\
columns do not reach their maximum values until $\sim10^6$\,yr, but
the other high ions peak in abundance at earlier interface age.
The theory predicts
$N$(\os)~$=10^{12}-10^{13}$\sqcm\ per interface (depending on the
magnetic field direction), which we can correct by a factor of 0.8 to
account for non-solar abundances. Even if we assume a magnetic field
normal to the cloud boundary, we would still require between six to eighteen
interfaces (i.e., three to nine clouds), to account for the observed
\os\ column 
densities in Complex~C of $10^{13.66}-10^{14.12}$, with a larger
number if the magnetic field has a component parallel to the boundary,
quenching the conduction. In one particular case, we already know of a
multiple component structure: two high-velocity neutral components are
thought to exist toward \pga\ \citep{Ri01a, Se03b}, providing
four interfaces. Therefore, at least one more low density neutral
absorber at Complex~C velocities, with a pair of hot interfaces,
is necessary to explain the \os\ toward \pga\ in the CI model. Such a
component could be too close in velocity to the $-128$\kms\ component
to be resolved with current instruments.

The $N$(\sif) prediction may be the least robust of the various
models, since \sif\ has the lowest creation ionization potential
(33.5\,eV) of any of the high ions under study, and so is 
most susceptible to production by photoionization, for example  
by photons escaping the Galactic disk.
This might lead to enhanced $N$(\sif)/$N$(\os) ratios over the individual
model predictions, raising all boxes vertically upward toward the
\pga\ data point on Figure 7. Photoionization could also enhance the
$N$(\cf)/$N$(\os) ratio, if the radiation field were hard enough.
On the other hand, depending on the slope of the ionizing spectrum, 
the $N$(\sif)/$N$(\os) ratio could also {\it decrease} if enough
\ion{Si}{5} is formed to depress the \sif\ prediction. A more detailed
understanding of the escaping radiation filed is needed to fully
investigate the effect of additional photoionization.
If it does play a role, then multiple ionization processes
are occuring in HVCs, just as multiple ionization processes appear to
be at work in the hot Galactic halo \citep{II88, Sh94, Sa03, Zs03}. 
 
\subsection{Kinematic Information}

Unfortunately, given the paucity of data points and the uncertainty in
the model predictions stemming from the numerous metallicity issues,
we cannot at this point uniquely specify the ionization mechanism
purely from the ionic ratios. Analysis of ionic ratios has the further
problem that multiple, separate regions of absorbing gas, closely related in
velocity space, can contribute to the integrated column density along
a line of sight, making interpretation of ionic ratios very
difficult. However, the {\it kinematics} of the
absorption provide further information; by 
comparing the extent and line centers of high-velocity absorption between
species, one can gain useful information regarding the ionization mechanism.

In the case of Mrk~279 and Mrk~817, high-velocity absorption is
detected in low and high ionization species at very similar
velocities (Mrk~279: $\bar{v}_{\mathrm{O~VI}}=-133\pm6$\kms,
$\bar{v}_{\mathrm{H~I}}=-137$\kms; Mrk~817: 
$\bar{v}_{\mathrm{O~VI}}=-109\pm10$\kms, $\bar{v}_{\mathrm{H~I}}=-109$\kms).
In the other three sight lines the difference is less than 20\kms.
In Table 8 we list our measured central velocities of the
high-velocity \os, and compare them to both the centers of the \hi\
and \ion{Si}{2} components, and to earlier measures of the \os\ velocities from
S03. In our sample the average displacement
$<\bar{v}_{\mathrm{O~VI}}-\bar{v}_{\mathrm{H~I}}>=3\pm12$\kms, where
the error represents the standard deviation of the sample, whereas
the S03 results (that used nine Complex~C sight lines) have an average
displacement 
$<\bar{v}_{\mathrm{O~VI}}-\bar{v}_{\mathrm{H~I}}>=0\pm13$\kms. The
differences between our measurements and the S03 measurements of
$\bar{v}_{\mathrm{O~VI}}$ are due to different data reduction
pipelines (v2.1.6 versus v1.8.7), and in the case of \pgb, our inclusion of new data. 
These changes caused the choice of velocity integration limits to change
(when $v$ is measured using the moment of the optical depth
technique), or by our using the component fitting method when the line
center was more clearly defined. We also measure
$<\bar{v}_{\mathrm{O~VI}}-\bar{v}_{\mathrm{Si~II}}>=0\pm15$\kms. 
These small differences between the velocities of \os, \ion{Si}{2}, and \hi\
reveal a strong kinematic connection between the neutral
and highly ionized gas in Complex~C. 

The observed velocity alignments are consistent with,
and indeed suggestive of, an origin at some form of interface 
between cold/warm and hot regions of gas. There is no reason
why the ion \os\ should line up with \hi\ otherwise, since these two
species trace different interstellar phases and do not exist in the
same physical regions, unless the \os\ is frozen-in at a far lower
temperature than that at which it normally appears. We stress that the
observed alignment between \os\ and neutral species is 
not an artifact of the velocity calibration procedure, which relied
upon tying the neutral species (but not the \os) in the \fuse\
bandpass with \hi\ emission components. 

In an interface scenario, any neutral cloud,
\emph{regardless of its depth}, contributes two interfaces to the line
of sight. Thus one would not expect to see a good correlation between
$N$(\os) and $N$(\hi) along lines of sight sampling hot and cold regions of gas
separated by interfaces. Indeed, \citet{Sa03} found no such
correlation when studying the \os\ distribution along 102 sight lines
through the Galactic halo. The ability of interface geometries to explain 
kinematic correspondences between low- and high-ionization absorbing
species has been recognised before \citep{Co79, Zs03}. Recently, \citet{Ho03}
concluded that interfaces along the line of sight toward globular
cluster star vZ~1128 are a plausible explanation for the similar kinematics
observed in low- and high-ionization species in that direction.

Toward \pga, a displacement of $18\pm6$\kms\ is observed between the 
high-velocity \os\ and \hi\ centroids. However, it is
already suspected that there are two high-velocity neutral gas components along
this sight line, with the $-128$\kms\ component containing enough
neutral gas to be seen in \hi\ emission, and the $-110$\kms\
component not detected in \hi\ emission, but seen in the more sensitive
\ion{O}{1} absorption line \citep{Se03b}. Interestingly, if we assume
the high-velocity \os\ toward \pga\ to be predominantly associated with the
$-110$\kms\ neutral gas component, so that
$\bar{v}_{\mathrm{O~VI}}-\bar{v}_{\mathrm{H~I}}=0\pm6$\kms, then the average 
displacement between neutral and highly ionized Complex~C gas becomes
$<\bar{v}_{\mathrm{O~VI}}-\bar{v}_{\mathrm{H~I}}>=-1\pm8$\kms. Such
a scenario, with two components having very different $N$(\os)/$N$(\hi) 
ratios, can easily be explained in the interface picture by two
neutral clouds of different column density, each with its own pair of
interfaces. Therefore the type of
displacement seen toward \pga\ is not inconsistent with the
interface theory, but rather can be explained by multiple interfaces. 
Table 9 contains a detailed study of all
the absorption species toward \pga\ for which we could measure a
precise line center to the Complex~C gas, using Gaussian component
fitting. Each of the species appears to be centered either
near $-128$\kms\ (designated Component 1), or near $-110$\kms\
(Component 2). Component 2 has a overall higher degree of ionization,
being seen in \cf\ and \os. Though assigning a component to each
absorption line is difficult on the basis of these velocities alone,
the stucture seen in Figure 4 and the fact that these components have
been reported elsewhere lend support to our conclusion that multiple
components exist toward \pga. 

The conductive interface (CI, \S6.4) and the turbulent mixing layer models
(TML, \S6.5) both have the attractive feature of naturally explaining the
closely aligned kinematics observed between low and high ionization
species. Recent studies have shown that
turbulent magnetic fields do not completely suppress thermal conduction
across interfaces \citep{Ch03}, so the true nature of the ionization
at HVC surfaces likely includes both turbulent and conduction
effects. Regardless of whether the energy is
transported by thermal conduction or turbulent mixing, our kinematic
observations are consistent with the core-interface structure of HVCs
suggested by \citet{FF94}, and discussed in \citet{Wo95}. 

Complex~C is known to have an \hi\ component at velocities
$-205\le v_{LSR}\le-155$\kms, along a ridge passing spatially along the middle
of the main cloud \citep{Wa01}. This feature can be seen in Figure 1,
and was dubbed the ``high-velocity ridge'' (HVR)
by \citet{Tr03}. These authors found similar morphologies and low-ion
column density ratios in the HVR and Complex~C proper,
suggesting similar abundances and physical conditions in the two parts
of the complex. Of the five sight lines we study in detail here,
Mrk~876 and \pgb\ pass through the HVR, with Mrk~279 just adjacent to the
$N$(\hi)~$=5\times10^{18}$\sqcm\ contour, whereas
Mrk~817 and \pga\ clearly contain no HVR \hi\ emission. Interestingly,
as seen in Table 4, Mrk~817 and \pga\ are the same two sight lines
whose velocity limits of \os\ absorption only extend to $-160$\kms. In
other words, those sight 
lines that display HVR \hi\ emission also display HVR \os\
absorption. This is further evidence of a kinematic correspondence
between the low and high ionization species.

Analysis of the widths of the high-velocity components can
give information on the temperature in the absorbing gas. Fully interpreting
line widths is difficult because of line blending, the presence of
multiple components, instrumental broadening, and non-thermal
broadening. However, we find that our observed high-velocity \os\ line widths
lie in the range $b=27-52$\kms\ (not accounting for instrumental
broadening, which for \fuse\ is of order $15-20$\kms). A
purely thermally broadened \os\ absorber at 300,000\,K would have
$b=18$\kms, suggesting that some non-thermal process is contributing
to the line widths. However, it is also possible that these detections
are multiple 
thermally-broadened components blended together, especially toward
\pga\ where we already know of two high-velocity components. One explicit
prediction of the conductive interface theory is that the \os\ lines
should be predominantly thermally broadened -- unfortunately, there
are too many unknowns to fully test this prediction, but our data are not
inconsistent with it.

S03 noticed a trend in which the value of $N$(\os)/$N$(\hi) in Complex
C increases toward lower longitudes and latitudes. \citet{Tr03}
interpreted this as evidence supporting their claim that the lower latitude
and lower longitude parts of the complex are interacting more
vigorously with the surrounding medium, forming a ``leading edge'' of
the complex. This claim is based on their detailed study of the
physical conditions in the gas towards 3C~351. With
our small sample of five sight lines with new measurements of \os, it 
is difficult to either confirm or refute this claim. However, we note
that the variation in $N$(\os)/$N$(\hi) is predominantly caused by
variation in $N$(\hi), which changes by a factor of $\approx20$ over
the face of Complex~C, whereas $N$(\os) only varies by a factor of $\approx3$.
In the conductive interface scenario that we favor, this situation
could reflect a cloud geometry in which 
the lower longitude and latitude parts of the cloud have a lower depth
of neutral gas, but still have a similar number of interfaces and hence
similar $N$(\os).

By using the \os\ a tracer of a hot shell of material surrounding
Complex~C, one can estimate the total amount of mass contained in the
hot ionized gas phase of the complex, according to 
\begin{eqnarray}
M_{\mathrm{H^+(hot)}}&=&\frac{M_{\mathrm{H^+(hot)}}}{M_{\mathrm{H~I}}}\;.\;M_{\mathrm{H~I}}=\frac{N_{\mathrm{H^+(hot)}}}{N_{\mathrm{H~I}}}\;.\;M_{\mathrm{H~I}}\nonumber\\ 
                     &=&\frac{N_{\mathrm{O~VI}}f_{\mathrm{O~VI}}^{-1}(\mathrm{O/H})^{-1}_{\odot}Z^{-1}}{N_{\mathrm{H~I}}}\;.\;M_{\mathrm{H~I}}
\end{eqnarray}

\noindent where $f_{\mathrm{O~VI}}$ is the fraction of oxygen atoms in
the five-times-ionized state, and
$(\mathrm{O/H})_{\odot}Z=A_{\mathrm{O}}=10^{-4.31}$ is the oxygen
abundance in Complex~C. Using the average observed ratio of 
$N_{\mathrm{O~VI}}/N_{\mathrm{H~I}}=1.8\times10^{-6}$ in Complex~C
(S03) and
$f_{\mathrm{O~VI}}<0.2$, we find the mass of Complex~C in
\os-bearing gas to be approximately 18\% of the
\hi\ mass, or $\approx15$\% of the total (\hi+\ion{H}{2}) hydrogen
mass. This fraction would decrease if a significant amount of \ion{H}{2}
is contained within the warm ($T\approx8000$\,K) ionized
material, but would increase if $f_{\mathrm{O~VI}}\ll0.2$. \citet{Wa01} has 
estimated the neutral mass of Complex~C to be 
$M_{\mathrm{H~I}}=3\times10^6(d/\mathrm{6\,kpc})^2M_{\odot}$, so that
$M_{\mathrm{H^+(hot)}}\approx5\times10^5\,M_{\odot}$ if the distance
to Complex~C is 6\,kpc.

In the interface scenario, the detection of highly 
ionized gas in HVCs offers further evidence (albeit indirect) for the
presence of a hot extended corona or Local Group Medium with which Complex~C is
interacting. Several lines of evidence already exist that indicate a
interaction between HVCs and the low density gaseous
Galactic halo \citep{Be99, QM01}. These include (a) the pattern of
decreasing cloud velocities for clouds closer to the Galactic 
plane, expected if drag dominates the cloud infall, (b) cometary
(head/tail) HVCs and evidence of stripping, (c) H$\alpha$ enhancement on 
cloud edges, (d) X-ray emission possibly associated with Complex~C,
and (e) high non-thermal \hi\ pressures. Studies with the {\it Chandra}
and {\it XMM-Newton} X-ray satellites have recently detected
both \ion{O}{7} and \ion{O}{8} absorption near zero velocity
\citep{Fa03, Ni03, Ra03}. These ions could be tracing the same
extended corona that we believe is interacting with Complex~C.

\section{Insights from Relevant Nearby Sight Lines}

Several sight lines passing close to Complex~C have been observed with
\fuse\ - we briefly mention the important properties of the most
relevant, since it is of interest how highly ionized gas in these
directions may relate to Complex~C. 
 
\subsection{H1821+643}

\citet{Tr03} have published absorption line measurements of the sight
line toward H1821+643, passing near Complex~C. There is a
highly ionized absorption component in this direction at $-212$\kms,
with unusual ionization properties: seen in \cf\ and \os, but
not in \hi\ emission or \ion{Si}{3} absorption. H1821+643
is a direction that passes through the Outer Arm of the Milky Way, but
the high-velocity component is well separated from the Outer Arm absorption.
Using the \citet{Tr03} measurements of \sif, \cf, and \nf\ together
with the S03 measurement of \os\ we find
$N$(\sif)/$N$(\os)~$<0.02$, $N$(\cf)/$N$(\os)~$=0.32\pm0.13$, and  
$N$(\nf)/$N$(\os)~$<0.10$, similar to what is observed in Complex
C. However, it is not clear whether this component has any affiliation
with Complex~C, since the absorption is centered almost 100\kms\ away
from the highly ionized gas seen in Complex~C proper. Furthermore, the
H1821+643 sight line passes more than $5\degr$ from where gas at such
velocities is seen in \hi. The \hi\ Lyman series 923.150, 920.963, and
919.351\,\AA\ lines in the \fuse\ spectrum show no evidence for a
component at $-212$\kms, setting a $3\sigma$ upper limit of
$\approx80$\,m\AA\ for a feature assumed to be 25\kms\ wide,
corresponding to $N$(\hi)~$<10^{16}$\sqcm.

\subsection{Mrk~290, Mrk~487, and 3C~249.1}

Mrk~290, found at $l=91.49\degr$, $b=47.95\degr$, passes through
Complex~C but was rejected for study in this paper because of its low
S/N. However, 
high-velocity \os\ and \hi\ are both detected in Complex~C in this direction
\citep{Wa03}. Mrk~487, lying near Mrk~290 at $l=87.84\degr$, $b=49.03\degr$
but outside the contour delineating the \hi\ edge of Complex~C, shows {\it
neither} high-velocity \hi\ or \os, although there is an IVC (core
IV15) seen at $-85$\kms. The 3$\sigma$ limit for high-velocity \os\
toward Mrk~487 is $N$(\os)$\,<14.13$, unfortunately not a strong
constraint. An \os\ non-detection would represent an important 
result, since in this case the interface surrounding Complex~C would
have to be thin enough not to be detected toward Mrk~487. 3C~249.1 is
another sight line a few degrees off the edge of Complex~C, in
which neither high-velocity \hi\ or \os\ is seen. \citet{Bo90} suggest
a typical conductive interface thickness is 15\,pc, which corresponds
to 5\arcmin\ if
one assumes the distance to Complex~C is 10\,kpc, so indeed we would
expect a sharp cut-off to the interface when looking off the side of
the complex.

\subsection{NGC~5447 and NGC~5471}

NGC~5447 and NGC~5471 are two \ion{H}{2} regions in the spiral galaxy
M101, the sight line to which passes through a hole in Complex
C. NGC~5447 has good \fuse\ data showing high-velocity \os\ absorption, but no 
high-velocity \hi\ is present in the Leiden-Dwingeloo Survey
\citep{HB97}. However, 
this is probably due a low column density of neutral gas, since the
sensitive \ion{C}{2} line shows extended absorption out to
$\approx-200$\kms, suggesting $N$(\hi)~$\sim10^{17}$\sqcm. NGC~5471 has an
\os\ tail extending to $-150$\kms, and no high-velocity \hi, but the
\ion{C}{2} only extends to $-100$\kms. Difficult
continua make precise measurements difficult in these directions, but
the high-velocity \os\ detections could be sampling the Complex~C interface.

\section{Conclusions}

We have investigated the properties of high ion absorption associated
with HVC Complex~C and its relation to low ion absorption and emission
using \fuse\ and \hst\ absorption line spectroscopy. We summarize the 
results of our study in the following key points, numbers 1, 3, 5, and
7 of which verify the basic findings of S03: 

\begin{enumerate}

\item{In all \fuse\ sight lines through Complex~C where \hi\ emission has
  been detected, we observe high-velocity \os\ absorption . Of the
  five Complex~C sight lines showing high-velocity absorption studied
  here, the mean logarithmic column density is
  $\langle\mathrm{log}~N\rangle=13.82$, with a standard deviation of
  0.21 dex.}  

\item{High-velocity \nf\ absorption is detected at $3.2\sigma$
  significance in one Complex~C sight line (Mrk~279). The
  non-detection of high-velocity \nf\ toward the other sight lines is
  consistent with the low N/O abundance ratio previously measured in
  the neutral gas of Complex~C.} 

\item{We find that in all five Complex~C sight lines, the \hi\ and
  \os\ high-velocity components are centered within 20\kms\ of one
  another, with an average displacement of 
  $<\bar{v}_{\mathrm{O~VI}}-\bar{v}_{\mathrm{H~I}}>=3\pm12$\kms; we also
  measure $<\bar{v}_{\mathrm{O~VI}}-\bar{v}_{\mathrm{Si~II}}>=0\pm15$\kms. In
  the directions along the high-velocity ridge where the \hi\ emission
  extends down to $-200$\kms, so does the \os\ absorption, indicating a close
  kinematic correspondence between neutral and highly ionized gas.}

\item{We measure high ion column density ratios in the high-velocity
  Complex~C gas. Along the \pga\ sight line,
  $N$(\sif)/$N$(\os)~$=0.10\pm0.02$,
  $N$(\cf)/$N$(\os)~$=0.35^{+0.05}_{-0.06}$, and
  $N$(\nf)/$N$(\os)~$<0.07$. The $N$(\nf)/$N$(\os) ratio is
  $0.19^{+0.06}_{-0.07}$ toward Mrk~279, $<0.11$ toward Mrk~876, and
  $<0.35$ for the \pgb\ sight line (all upper limits are 3$\sigma$).}

\item{Collisional ionization equilibrium and photoionization by
  the extragalactic radiation field can be ruled out as the origin of
  the highly ionized gas in Complex~C. Our observed
  $N$(\sif)/$N$(\os), $N$(\cf)/$N$(\os), and $N$(\nf)/$N$(\os) ionic ratios are
  most consistent with the conductive interface and turbulent mixing
  layer models. The shock ionization and radiative cooling models are
  unable to simultaneously reproduce these ratios.}

\item{We consider it likely that the \os\ observed at HVC velocities
  is produced in conductive or turbulent interfaces at the boundaries of
  Complex~C. The ionic column densities and their ratios between the
  highly ionized species, the coincidence in central velocity of low,
  (intermediate), and high ion absorption components, and the similar
  velocity extent of \hi\ emission and \os\ absorption all lend support 
  to this hypothesis. More \hst/STIS observations of Complex~C sight
  lines would be crucial
  in discriminating between the CI and TML models; measurement of the 
  $N$(\cf)/$N$(\os) ratio would be the crucial diagnostic test} 

\item{The interface hypothesis, if correct, provides indirect evidence
  for the existence of a hot, low-density medium surrounding and
  interacting with Complex~C. This medium would take the form of an
  extended Galactic corona or a diffuse intergroup medium, depending
  on the location of Complex~C.} 

\item{We suggest an approximate method for scaling column density
  ratio predictions to low metallicity regions. However, there is
  considerable need for the ionic ratio predictions of 
  several ionization mechanism models to be updated to include new
  solar abundance measurements and applicability to low-metallicity
  environments.}  

\end{enumerate}

The STIS observations of \pga\ were obtained through \hst\ program 8695,
with financial support from NASA Grant GO-08695.01-A from the Space
Telescope Science Institute. US participants appreciate financial
support from NASA contract NAS5-32985. B. P. W. acknowledges support
by NASA grant NAG5-9179. P. R. is supported by the {\it Deutsche
Forschungsgemeinschaft}. T. M. T. appreciates support from NASA Long Term
Space Astrophysics grant NAG5-11136.

\begin{deluxetable}{lcccc ccccc r}
\tablewidth{0pt}
\tabcolsep=4pt
\tabletypesize{\footnotesize}
\tablehead{
Target & $l$ & $b$ & $z$ & Type\tablenotemark{a} &
  $B$\tablenotemark{b} & $E(B-V)$\tablenotemark{c} &
  log $N$(\hi)\tablenotemark{d} & $\Delta
  v_{LSR}$\tablenotemark{e} &
  [\ion{O}{1}/\hi]\tablenotemark{f} &
  [\ion{N}{1}/\hi]\tablenotemark{f}\\
& ($\degr$) & ($\degr$) & & & & & ($N$ in cm$^{-2}$) & (km~s$^{-1}$) & &}
\tablecaption{Sight Line Data}
\startdata
Mrk~279 & 115.04 & 46.86 & 0.0304 & Sey1 & 15.15 & 0.016 &
19.49$^{+0.06}_{-0.08}$ & +12.2 & $-0.71^{+0.36}_{-0.25}$ & $<-1.20$\\
Mrk~817 & 100.30 & 53.48 & 0.0315 & Sey1 & 14.19 & 0.007 &$19.49\pm0.01$ & +14.1 & $-0.59^{+0.25}_{-0.17}$ & $<-1.35$\\
Mrk~876 & \phn98.27 & 40.38 & 0.1289 & Sey1 & 16.03 & 0.027 &
$19.28\pm0.02$ & +15.4 & \nodata & $-1.09^{+0.16}_{-0.15}$\\
\pga\ & 120.56 & 58.05 & 0.4778 & QSO & 15.84 & 0.008 &
$19.94\pm0.06$\tablenotemark{g} & +10.8 & $-0.79^{+0.12}_{-0.16}$\tablenotemark{g} &
$-1.82^{+0.20}_{-0.13}$\\ 
\pgb\ & 111.89 & 52.02 & 0.0882 & Sey1 & 14.54 & 0.020 &
$19.75^{+0.01}_{-0.02}$ & +12.5 & $<-0.79$ & $<-0.88$\\
\enddata
\tablenotetext{a}{Target type taken from NASA/IPAC Extragalactic
  Database (NED), available online at {\tt http://nedwww.ipac.caltech.edu}.}
\tablenotetext{b}{Apparent blue magnitude taken from the catalog of
  \citet{VV00}.}
\tablenotetext{c}{Galactic foreground interstellar reddening, based on
  measurements by \citet{Sc98}.}
\tablenotetext{d}{Neutral hydrogen column density in Complex~C from
  \citet{Wa01}.} 
\tablenotetext{e}{Velocity correction from heliocentric to LSR
  reference frames, given by $\Delta v_{LSR}=v_{LSR}-v_{Helio}$, 
  assuming the IAU convention \citep{KL86} that the Sun is moving at
  20\kms\ in the direction $18^{\mathrm{h}},+30\degr$ (J1900),
  i.e. $l=56.16\degr$, $b=22.76\degr$. This definition of
  the Standard Solar Motion is adopted throughout this paper.} 
\tablenotetext{f}{Metallicity measurements in Complex~C gas published
  by \citet{Co03a}. All upper limits are 3$\sigma$.}
\tablenotetext{g}{Exception: Complex C $N$(\hi) and oxygen abundance
  toward \pga\ from \citet{Se03b}.}
\end{deluxetable} 

\begin{deluxetable}{lcccr}
\tablewidth{0pt}
\tablehead{
Sight Line & Program ID & Channel & Date & $t_{exp}$\\
& & & (yr-mn-dy) & (ks)}
\tablecaption{\fuse\ Observations of Complex~C\tablenotemark{a}}
\startdata
Mrk~279 & P1080303 & LiF1A/LiF2B & 1999-12-28 & 48.9 \\
	& P1080304 & LiF1A/LiF2B &2000-01-11 & 30.7 \\
Mrk~817 & P1080403 & LiF1A & 2000-12-23 & 76.3 \\
	& P1080404 & LiF1A & 2001-01-18 & 86.0 \\
Mrk~876 & P1073101 & LiF1A/LiF2B & 1999-10-16 & 52.8 \\ 
\pga\ & P1080101 & LiF1A & 2000-02-25 & 52.4 \\
	    & P1080102 & LiF1A & 2000-12-25 & 57.9 \\
    	    & P1080103 & LiF1A & 2001-01-29 & 82.2 \\	
   	    & P1080104 & LiF1A & 2001-03-12 & 106.4 \\	
   	    & P1080105 & LiF1A & 2001-03-14 & 105.0 \\	
   	    & P1080106 & LiF1A & 2001-03-17 & 100.5 \\	   	 
            & P1080107 & LiF1A & 2001-03-19 & 96.8 \\	
   	    & P1080108 & LiF1A & 2001-03-22 & 33.7 \\	
   	    & P1080109 & LiF1A & 2001-03-28 & 33.6 \\	
\pgb\ & P1072501 & LiF1A/LiF2B & 2000-01-18 & 70.2 \\
	    & S6010701 & LiF1A/LiF2B & 2002-02-01 & 50.5 
\enddata
\tablenotetext{a}{All \fuse\ observations have resolution (FWHM)
  $\approx20$\kms, pixel size 2.1\kms, were taken with the $30\arcsec\times30\arcsec$ 
  aperture, and have been rebinned by five pixels for display in this paper.}
\end{deluxetable} 

\begin{deluxetable}{lcccc ccccc}
\tablewidth{0pt}
\tabcolsep=4pt
\tabletypesize{\footnotesize}
\tablehead{
Sight Line & Inst. & Grating & Dataset & Date & $t_{exp}$ &
  Wavelength & FWHM\tablenotemark{a} & Pixel Size & Rebinning\tablenotemark{b}\\
  & & & & (yr-mn-dy)& (ks) & (\AA) & (km~s$^{-1}$) & (km~s$^{-1}$) & \\
  (1) & (2) & (3) & (4) & (5) & (6) & (7) & (8) & (9) & (10)}
\tablecaption{\hst\ Observations of Complex~C}
\startdata
Mrk~279 & GHRS & G140M & Z3E7030$x$T  & 1997-01-16 & $3\!\times\!6.1$
& 1223--1258 & 19 & 4.4 & 2\\
& & & $x$=$4,6,8$ & & & & & & \\
& &	       & Z3E7030AT & 1997-01-16 & 1.5 & & & &\\	
Mrk~817 & GHRS & G160M & Z3E70104T & 1997-01-12 & 9.5 & 1223--1258 &
19 & 4.4 & 2\\
&	&      & Z3E70106T & 1997-01-12 & 8.8 & & & &\\
&	&      & Z3E70108T & 1997-01-12 & 8.5 & & & &\\
Mrk~876 & STIS & G140M & O4N308010 & 1998-09-19 & 2.3 & 1194--1249 &
30 & 12.9\phn & 1\\
\pga\ & STIS & E140M & O63G0$x$0$y$0 & 2001-01-17 & $30\!\times\!2.4$ &
1142--1730 & 9 & 3.4 & 3\\
& & & $x$=$5,6,7,8,9$ & 2000-01-18 & & & & &\\
& & & $y$=$1,2,3,4,5,6$ & 2000-01-19 & & & & &\\
& & & O63G110$x$0 & 2001-01-19 & $\phn4\!\times\!2.3$ & & & &\\
& & &  $x$=$1,2,3,4$ & & & & & &\\
% & & & O4N307020 & 2001-12-19 & 1.8 & & & &\\
\pgb\ & STIS & G140M & O4EC54010 & 2000-08-02 & 8.5 & 1194--1248
& 30 & 6.4 & 2\\
& &	       & O4EC54020 & 2000-08-02 & 6.3 & & & & 
\enddata
\tablenotetext{a}{Instrumental resolution, specifically the full-width
  at half-maximum (FWHM) of the line spread function, according to the
  GHRS \citep{So95} and STIS \citep{Pr02} instrument handbooks.}
\tablenotetext{b}{The rebinning factor applied to produce bins of $\approx10$\kms\
  width, used for consistency in our plots comparing \fuse\ and \hst\ spectra.}
\end{deluxetable} 

\begin{deluxetable}{llccc ccccr}
\tablewidth{0pt}
\tabcolsep=4pt
\tabletypesize{\footnotesize}
\tablehead{
Sight Line & Ion & $\lambda$\tablenotemark{a} & $v_{min,max}$ & $\bar{v}$ & 
  $b$ & Method\tablenotemark{b} & $W_{\lambda}$\tablenotemark{c} &
  log\,$N_a$\tablenotemark{d} & S/N\tablenotemark{e}\\  
  & & (\AA) & (km~s$^{-1}$) & (km~s$^{-1}$) & (km~s$^{-1}$) &  & (m\AA)
  & ($N_a$ in cm$^{-2}$) &}
\tablecaption{Measurements of High Ions in Complex~C}
\startdata
Mrk~279 & \ion{S}{3} & 1012.495 & $-210,-115$ &
\nodata\tablenotemark{f} & \nodata\tablenotemark{f} & \nodata\tablenotemark{f} &
$0\!\pm\!6\!\pm\!9$\tablenotemark{f} & $<13.67$ (3$\sigma)$ & 27\\
& \nf & 1238.821 & $-190,-115$ & $-149\!\pm\!5$ & $41\!\pm\!7$ & CF & 
$17\!\pm\!5\!\pm\!3$\phn &
$12.93^{+0.12\;+0.02}_{-0.17\;-0.01}$ & 33\\  
& \os\ & 1031.926 & $-210,-115$ & $-133\!\pm\!6$ & $52\!\pm\!5$ & CF &
$53\!\pm\!6\!\pm\!7$\phn &$13.66^{+0.04\;+0.05}_{-0.04\;-0.04}$ & 31\\

Mrk~817 & \ion{S}{3} & 1012.495 & $-160,-80\phn$ & $-114\!\pm\!9$ &
$33\!\pm\!5$ & MOD & $32\!\pm\!3\!\pm\!10$ &
$13.95^{+0.04\;+0.13}_{-0.05\;-0.18}$ & 28\\  
	& \os\ & 1031.926 & $-160,-80\phn$ & \phn$-109\!\pm\!10$ & 
$29\!\pm\!6$ & MOD & $93\!\pm\!2\!\pm\!29$ &
$13.97^{+0.02\;+0.08}_{-0.02\;-0.11}$ & 35\\ 

Mrk~876 & \nf & 1238.821 & $-220,-100$ &
\nodata\tablenotemark{f} & \nodata\tablenotemark{f} & \nodata\tablenotemark{f} & 
\phn$4\!\pm\!17\!\pm\!9$\tablenotemark{f} & $<13.43$ (3$\sigma$) & 16\\
	& \os\ & 1031.926 & $-220,-100$ & $-148\!\pm\!9$ &
 $42\!\pm\!5$ & MOD & $132\!\pm\!6\!\pm\!41$\phn &
	$14.12^{+0.02\;+0.09}_{-0.02\;-0.11}$ & 18\\ 

\pga\ & \ion{S}{3} & 1012.495 & $-160,-80\phn$ & $-104\!\pm\!9$ &
$22\!\pm\!6$ & MOD & $17\!\pm\!5\!\pm\!12$ &
$13.69^{+0.14\;+0.18}_{-0.23\;-0.23}$ & 22\\  
& \sif & 1393.755 & $-160,-80\phn$ & $-119\!\pm\!3$ & $22\!\pm\!3$ & CF &
 $41\!\pm\!3\!\pm\!7$\phn & $12.73^{+0.02\;+0.03}_{-0.01\;-0.02}$ & 17\\
& \cf & 1548.195 & $-160,-80\phn$ & $-106\!\pm\!4$ & $28\!\pm\!4$ & CF & 
$60\!\pm\!5\!\pm\!3$\phn & $13.26^{+0.03\;+0.01}_{-0.04\;-0.02}$ & 9\\
& \nf & 1238.821 & $-160,-80\phn$ &
\nodata\tablenotemark{f} & \nodata\tablenotemark{f} & \nodata\tablenotemark{f} &
$2\!\pm\!5\!\pm\!2$\tablenotemark{f} & $<12.85$ (3$\sigma$) & 12\\
& \os\ & 1031.926 & $-160,-80\phn$ & $-110\!\pm\!5$ & $35\!\pm\!2$ & CF &
$57\!\pm\!5\!\pm\!17$ & $13.71^{+0.04\;+0.05}_{-0.04\;-0.05}$ & 34\\

\pgb\ & \nf\ & 1238.821 & $-190,-120$ & 
\nodata\tablenotemark{f} & \nodata\tablenotemark{f} & \nodata\tablenotemark{f} & 
\phn$0\!\pm\!9\!\pm\!15$\tablenotemark{f} & $<13.39$ (3$\sigma$) & 23\\ 
& \os\ & 1031.926 & $-190,-120$ & \phn$-147\!\pm\!10$ & $27\!\pm\!6$ & MOD
& $49\!\pm\!8\!\pm\!14$ & $13.66^{+0.08\;+0.11}_{-0.08\;-0.15}$ & 21
 
\enddata
\tablenotetext{a}{Rest vacuum wavelengths are taken from \citet{Mo91},
  as are the oscillator strengths (not quoted here).}
\tablenotetext{b}{Method for measuring velocity centroid and line
  width: CF = component fit, MOD = moment of optical depth (see text).}
\tablenotetext{c}{Equivalent width of absorption in
  the high-velocity gas, integrated between $v_{min}$ and $v_{max}$. The first
  error (statistical) accounts for photon noise statistical errors and
  continuum placement uncertainty. The second error (systematic)
  accounts for fixed pattern noise and choice of velocity
  integration limits.}
\tablenotetext{d}{Column density calculated using AOD technique \citep{SS91},
  including statistical plus continuum placement errors, and systematic
  errors, respectively.}
\tablenotetext{e}{Signal-to-noise ratio per resolution element in the
  combined spectrum near the spectral line in question.} 
\tablenotetext{f}{No significant detection.}
\end{deluxetable} 

\begin{deluxetable}{lcccc ccccc cc}
\tablewidth{0pt}
\tabcolsep=4pt
\tablehead{
Authors\tablenotemark{a} & \multicolumn{5}{c}{\underline{\phm{fillingfillin}
12 + log\,A$_{\mathrm{X}}^{\odot}$\phm{fillingfillin}}} 
& \multicolumn{5}{c}{\underline{\phm{fillingfillingsss}$\Delta$log\,A$_{\mathrm{X}}^{\odot}$\tablenotemark{b}\phm{fillingfillingsss}}}
& Used in \\
& {C} & {N} & {O} & {Si} & {S} & {C} & {N} & {O} & {Si} & {S} & Model\tablenotemark{c}}
\tablecaption{Solar Elemental Abundances}
\startdata
A73 & 8.52 & 7.96 & 8.82 & 7.52 & 7.20 & $-$0.13 & $-$0.03 & $-$0.13
&+0.02 & +0.06 & RC \\
G84 & 8.69 & 7.99 & 8.91 & 7.55 & 7.24\tablenotemark{d}& $-$0.30 &
$-$0.06 & $-$0.22 &$-$0.01 & +0.02 & CI, TML \\
AG89 & 8.56 & 8.05 & 8.93 & 7.54 & 7.27 & $-$0.17 & $-$0.12 &
$-$0.24&\phs0.00 & $-$0.01 & CIE, SI\\
GN93 & 8.55 & 7.97 & 8.87 &\nodata& \nodata & $-$0.16 & $-$0.04
&$-$0.18 & \nodata\ & \nodata & \nodata \\
GNS96 &8.55 & 7.97 & 8.87 &7.56\tablenotemark{d}&7.26\tablenotemark{d}
& $-$0.16 & $-$0.04 &$-$0.18 & $-$0.02 & \phs0.00 & \nodata\\
GS98 &8.52 & 7.92 & 8.83 & 7.56\tablenotemark{d}&7.26\tablenotemark{d}
 & $-$0.13 & +0.01 & $-$0.14 &$-$0.02 & \phs0.00 & \nodata\\
H01 & 8.59 & 7.93 & 8.74 & 7.54 & \nodata & $-$0.20 & \phs0.00 &
$-$0.05 & \phs0.00 & \nodata& \nodata\\ 
AP01,02 & 8.39 & \nodata & 8.69 & \nodata & \nodata & \phs0.00 & 
\nodata &\phs0.00 & \nodata & \nodata & \nodata\\
\tableline
Adopted  &8.39 & 7.93 & 8.69 & 7.54 & 7.26 & \nodata &\nodata &\nodata
& \nodata & \nodata & \nodata
\enddata
\tablenotetext{a}{A73: \citet{Al73}; G84: \citet{Gr84}; AG89: \citet{AG89};
  GN93: \citet{GN93}; GNS96: \citet{Gr96}; GS98: \citet{GS98}; H01: \citet{Ho01}; AP01,02: \citet{AP01, AP02}.}
\tablenotetext{b}{$\Delta$log\,A$_{\mathrm{X}}^{\odot}=$~log\,A$_{\mathrm{X}}^{\odot}$(adopted)~$-$~log\,A$_{\mathrm{X}}^{\odot}$,
  as used in equation 11.}
\tablenotetext{c}{These models assumed the given abundances in their
  calculations. See \S6.} 
\tablenotetext{d}{Average of photospheric and meteoritic value.}
\end{deluxetable} 

\begin{deluxetable}{lcccc c}
\tablewidth{0pt}
\tabcolsep=4pt
\tablecaption{Summary of Metallicity Measurements of Complex~C}
\tablehead{Study & Solar Abundances & Sight Line & Ratio & Value & Updated to\tablenotemark{a}}
\startdata
\citet{Wa99} & AG89 & Mrk~290 & [S/H] & $-1.05\pm0.12$ & $-1.04\pm0.12$\\
\citet{Gi01} & AG89 & Mrk~279 & [\ion{S}{2}/\hi] & $-0.36\pm0.18$ & $-0.35\pm0.18$\\
 & & Mrk~290 & [\ion{S}{2}/\hi] & $-1.10\pm0.06$ & $-1.09\pm0.06$ \\
 & & Mrk~817 & [\ion{S}{2}/\hi] & $-0.48\pm0.06$ & $-0.47\pm0.06$ \\
\citet{Co03a}& GS98 & Mrk~279
 & [\ion{S}{2}/\hi] & $-0.27^{+0.16}_{-0.20}$ & $-0.27^{+0.16}_{-0.20}$\\
 & & Mrk~290 & [\ion{S}{2}/\hi] & $-1.00^{+0.16}_{-0.15}$ & $-1.00^{+0.16}_{-0.15}$\\
 & & Mrk~817 & [\ion{S}{2}/\hi] & $-0.34\pm0.08$ & $-0.34\pm0.08$  \\
 & & \pga & [\ion{S}{2}/\hi] & $-0.74\pm0.13$ & $-0.74\pm0.13$  \\
\citet{Ri01a}& GN93 & \pga & [\ion{S}{2}/\hi] & $-0.85^{+0.12}_{-0.15}$ & $-0.84^{+0.12}_{-0.15}$\\
\tableline
\citet{Co03a} & GS98 & Mrk~279 &[\ion{O}{1}/\hi] & $-0.71^{+0.36}_{-0.25}$ & $-0.57^{+0.36}_{-0.25}$\\
& & Mrk~817 & [\ion{O}{1}/\hi] & $-0.59^{+0.25}_{-0.17}$ & $-0.45^{+0.25}_{-0.17}$\\
& & \pga & [\ion{O}{1}/\hi] & $-1.00^{+0.19}_{-0.25}$ & $-0.86^{+0.19}_{-0.25}$\\
\citet{Ri01a} & GN93 & \pga & [\ion{O}{1}/\hi] & $-1.03^{+0.37}_{-0.31}$& $-0.85^{+0.37}_{-0.31}$  \\
\citet{Se03b} & AP01,02 & \pga & [\ion{O}{1}/\hi] & $-0.79^{+0.12}_{-0.16}$ & $-0.79^{+0.12}_{-0.16}$ \\
\citet{Tr03} & H01  & 3C~351& [\ion{O}{1}/\hi] & $-0.76^{+0.23}_{-0.21}$ & $-0.71^{+0.23}_{-0.21}$  
\enddata
\tablenotetext{a}{Assuming the adopted abundances in the bottom row of
  Table 5. We have not fully accounted for errors in the new solar abundances.}
\end{deluxetable} 

\begin{deluxetable}{lcccc}
\tablewidth{0pt}
\tabcolsep=4pt
\tabletypesize{\small}
\tablecaption{Complex~C High Ion Column Density Ratios toward \pga\ --
  Observation versus Theory\tablenotemark{a}}
\tablehead{
Sight Line / Model & Abundances\tablenotemark{b} & $N$(\sif)/$N$(\os)
& $N$(\cf)/$N$(\os) & $N$(\nf)/$N$(\os)}  
\startdata
Mrk~279 & Complex~C & \nodata\ & \nodata\ & $0.19^{+0.06}_{-0.07}$\\
\pga\ & Complex~C & $0.10\pm0.02$ & $0.35^{+0.05}_{-0.06}$ & $<0.07$ (3$\sigma$)\\
\tableline
C. I. E. $\;(T=2.0-5.0\times10^5$ K) & Solar &
0.000--0.009 & 0.005--0.27\phn & 0.018--1.55\phn\\
C. I. E. $\;(T=2.0-5.0\times10^5$ K) & Complex~C &
0.000--0.006 & 0.002--0.12\phn & 0.002--0.14\phn\\
Radiative Cooling (RC) & Solar & 
0.001--0.013 & 0.036--0.17\phn & 0.053--0.090\\
Radiative Cooling (RC) & Complex~C & 
0.000--0.008 & 0.016--0.078 & 0.005--0.008\\
Conductive Interfaces (CI) & Solar &
0.003--0.058 & 0.042--0.93\phn & 0.065--0.51\phn\\
Conductive Interfaces (CI) & Complex~C & 
0.002--0.035 & 0.019--0.42\phn & 0.006--0.045\\
Turbulent Mixing Layers (TML) & Solar & 
0.087--1.29\phn & 1.05--6.92 & 0.17--0.58\\
Turbulent Mixing Layers (TML) & Complex~C &
0.052--0.78\phn & 0.47--3.09 & 0.015--0.050\\
Shock Ionization (SI) & Solar & 
0.002--0.18\phn & 0.016--1.04\phn & 0.031--0.051\\
Shock Ionization (SI) & Complex~C &
0.001--0.11\phn & 0.007--0.46\phn & 0.003--0.004\\
\enddata
\tablenotetext{a}{See \S6 for a description of the model parameters.}
\tablenotetext{b}{The model entries marked ``Solar'' have been corrected for
  updates to the solar abundance ratios since the models were
  published (see equation 11 and Table 5). The model entries marked
  ``Complex~C'' have had a 
  second correction applied, accounting for the non-solar abundance
  ratios in Complex~C (see equation 13 and Table 6).} 
\end{deluxetable} 

\begin{deluxetable}{lcccc ccc}
\tablewidth{0pt}
\tabletypesize{\small}
\tablehead{
 & & \multicolumn{2}{c}{\underline{Sembach et al. (2003a)}\tablenotemark{b}} &
 \multicolumn{4}{c}{\underline{
\phm{fillingfillingfillingfill}This Paper\phm{fillingfillingfillingfill}}} \\
 Sight Line & $\bar{v}_{\mathrm{H\;I}}$\tablenotemark{a}  &
 $\bar{v}_{\mathrm{O\;VI}}$ &
 $\bar{v}_{\mathrm{O\;VI}}-\bar{v}_{\mathrm{H\;I}}$ &
 $\bar{v}_{\mathrm{O\;VI}}$ &
 $\bar{v}_{\mathrm{O\;VI}}-\bar{v}_{\mathrm{H\;I}}$ &
 $\bar{v}_{\mathrm{Si\;II}}$\tablenotemark{c} &
 $\bar{v}_{\mathrm{O\;VI}}-\bar{v}_{\mathrm{Si\;II}}$\tablenotemark{d}\\
& (km~s$^{-1}$) & (km~s$^{-1}$) & (km~s$^{-1}$) & (km~s$^{-1}$) 
& (km~s$^{-1}$) & (km~s$^{-1}$) & (km~s$^{-1}$)}
\tablecaption{Kinematic Study of Highly Ionized Complex~C Gas}
\startdata
Mrk~279 & $-137$ & $-154\pm10$ & $-17\pm10$    & $-133\pm6$\phn & $+4\pm6$ & $-146\pm6$ & $+13\pm5$\\
Mrk~817 & $-109$ & $-116\pm10$ & \phn$-7\pm10$ & $-109\pm10$    & \phs\phn$0\pm10$ & \phn$-98\pm6$ & $-11\pm8$\\
Mrk~876 & $-133$ & $-142\pm10$ & \phn$-9\pm10$ & $-148\pm9$\phn & $-15\pm9$\phn       & $-128\pm6$ & $-20\pm8$\\
\pga\   & $-128$ & $-110\pm10$ & $+18\pm10$    & $-110\pm5$\phn & $+18\pm6$\tablenotemark{d}\phn & $-123\pm6$ & $+13\pm2$\\
\pgb\   & $-154$ & $-124\pm10$ & $+30\pm10$    & $-147\pm10$    & \phn$+7\pm10$ & $-150\pm6$ & \phn$+3\pm9$
\enddata
\tablenotetext{a}{\hi\ velocites from Effelsberg 100\,m telescope
  observations \citep{Wa01}. In cases with multiple high-velocity \hi\
  components, we list the strongest. The \hi\ velocity errors are very
  small ($\approx\pm1$\kms).} 
\tablenotetext{b}{S03 and \citet{Wa03} quote approximate 1$\sigma$
  errors of $\pm10$\kms\ on these velocity measurements.}  
\tablenotetext{c}{$\bar{v}_{\mathrm{Si\;II}}$ is measured by Gaussian
  component fitting of the \ion{Si}{2} $\lambda1020.699$ line.}
\tablenotetext{d}{Note there is a reported \ion{O}{1} component at
  $-110$\kms. If the \os\ is associated with this component,
  $\bar{v}_{\mathrm{O\;VI}}-\bar{v}_{\mathrm{H\;I}}$ toward \pga\ becomes $0\pm6$\kms.}
\end{deluxetable} 

\begin{deluxetable}{lccc}
\tablewidth{0pt}
\tablehead{Ion & Line & $\bar{v}_{LSR}$ & Component\tablenotemark{a}\\
& (\AA) & (km~s$^{-1}$) &}
\tablecaption{Detailed Kinematics of Complex~C Gas toward \pga}
\startdata
\ion{Ar}{1}& 1048.220 & $-127\pm5$ & 1\\
\ion{O}{1} & 1302.169 & $-131\pm5$ & 1\\
\ion{Si}{2}& 1020.699 & $-123\pm6$ & 1\\
\ion{Si}{2}& 1526.707 & $-121\pm5$ & 1\\
\ion{S}{3} & 1012.495 & $-104\pm9$ & \phn2?\\
\sif\      & 1393.755 & $-119\pm3$ & \phn2?\\
\cf\       & 1548.195 & $-106\pm4$ & 2\\
\os\       & 1031.926 & $-110\pm5$ & 2\\
\enddata
\tablenotetext{a}{Component 1 is at $-128$\kms, Component 2 is at
  $-112$\kms, as found by detailed component fitting of the \hi\ Lyman
  series absorption lines \citep{Se03b}.}
\end{deluxetable} 

\begin{figure}[ht]
\figurenum{1}
\epsscale{0.5}
\plotone{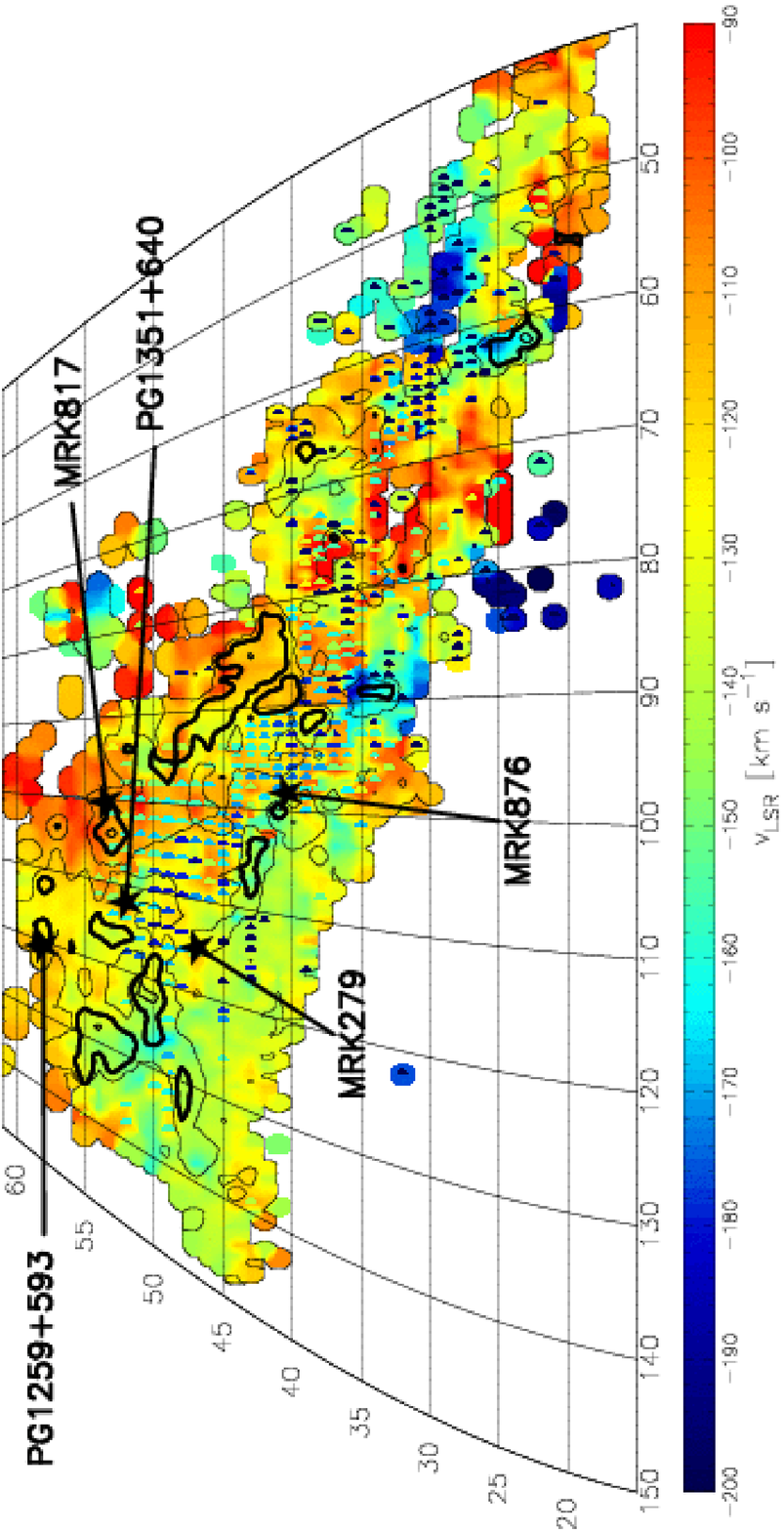}
\caption{\small \hi\ velocity field of Complex~C, over the range
  $-200<v_{LSR}<-80$\kms, from the Dwingeloo HVC Survey
  \citep{HW88}. The survey was  
  completed on a $1\degr\times1\degr$ grid with a $36\arcmin$ beam, 16
  \kms\ velocity resolution (FWHM) and a 0.05\,K detection limit. The
  sight lines we study in this paper are identified with stars. The
  contours represent column density levels of $0.2, 1.0$ and
  $3.0\times10^{19}$\sqcm. The half circles represent those directions
  with an additional higher velocity \hi\ component, dubbed the
  high-velocity ridge.} 
\end{figure}

\begin{figure}[ht]
\figurenum{2}
\epsscale{0.85}
\plotone{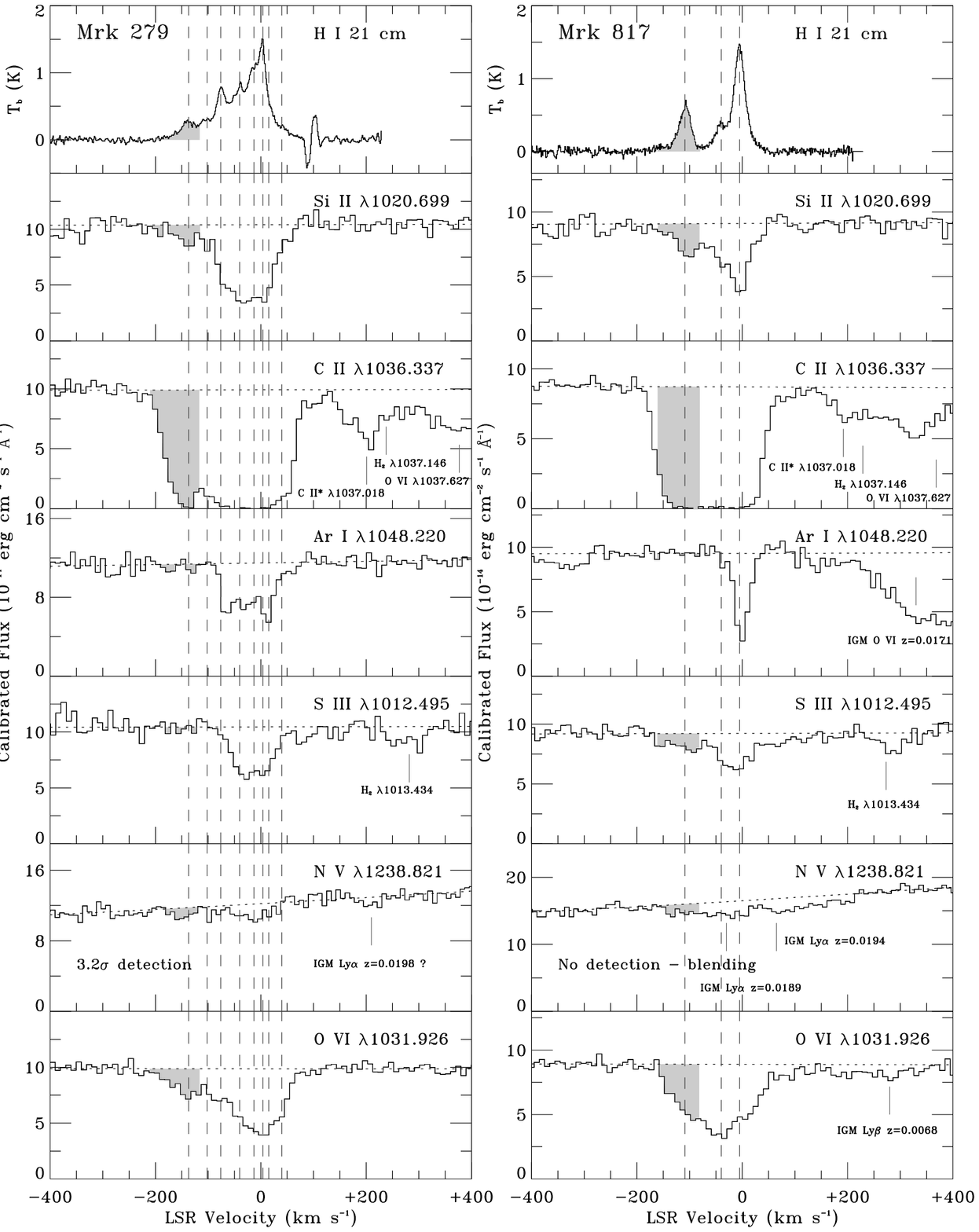}
\caption{\small \fuse\ absorption line profiles
  of \ion{Si}{2}, \ion{C}{2}, \ion{Ar}{1}, \ion{S}{3}, and \os, GHRS
  \nf\ data, and Effelsberg 100m radio telescope 
  \hi\ 21\,cm emission line observations \citep{WK01} toward Mrk~279 and
  Mrk~817. Gray shading is used to denote absorption (or emission) in
  the velocity range of Complex~C. Dashed lines indicate
  the peaks of the neutral hydrogen emission, and dotted lines indicate the
  position of our fitted continuum in each absorption line. Other
  absorption lines falling in the range $-400<v_{LSR}<400$\kms\ of each
  line are identified with annotated tick marks. High-velocity highly
  ionized gas is detected in
  \os\ along each sight line, and marginally detected in \nf\
  (3.2$\sigma$) toward Mrk~279. What appears to be high-velocity \nf\
  toward Mrk~817 is in fact a blend with intergalactic Ly$\alpha$ at
  $z=0.0189$.}
\end{figure}

\begin{figure}[ht]
\figurenum{3}
\epsscale{0.85}
\plotone{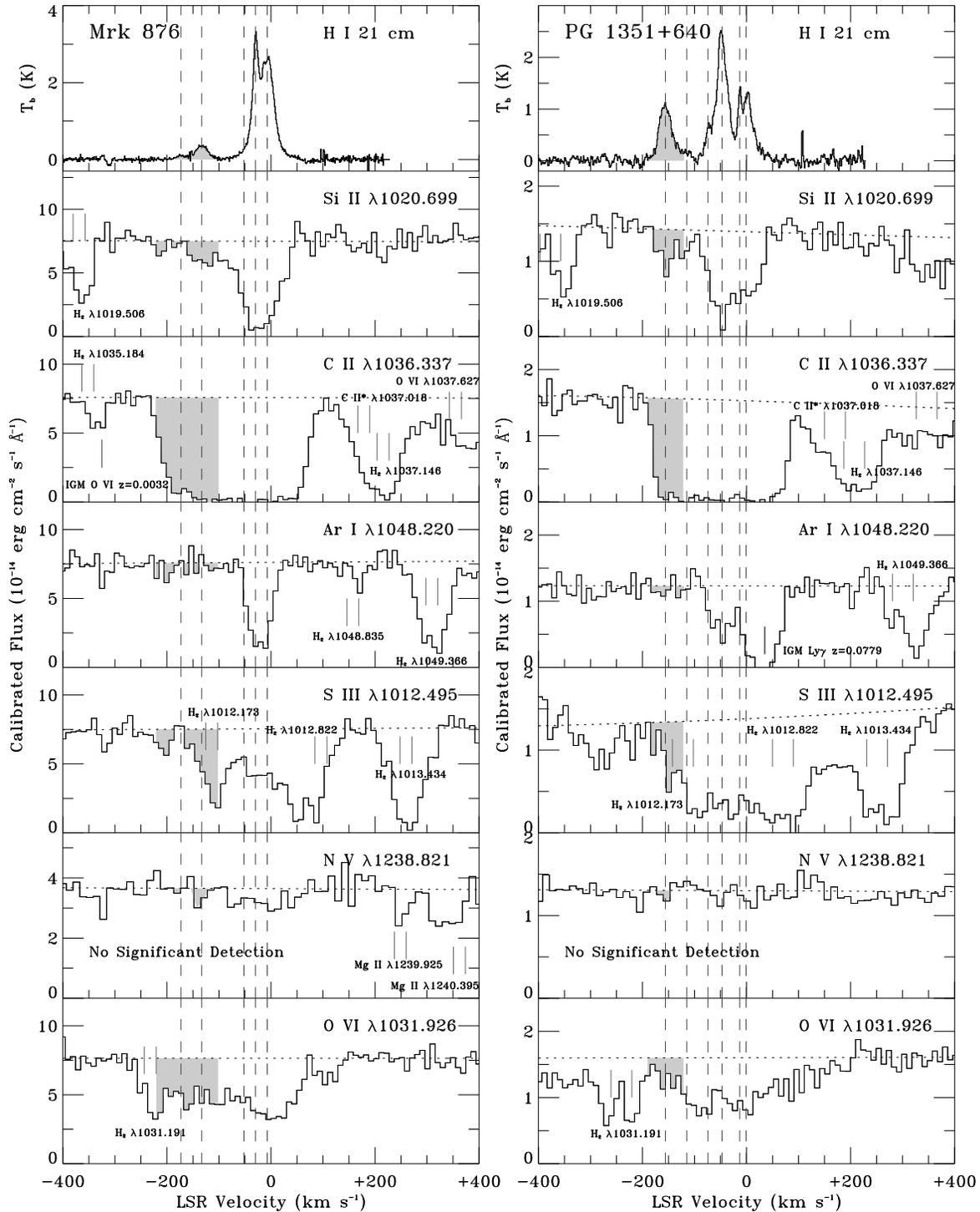}
\caption{\small \fuse\ absorption line profiles of \ion{Si}{2},
  \ion{C}{2}, \ion{Ar}{1}, \ion{S}{3}, and \os, STIS \nf\ data, and
  Effelsberg 100m radio telescope  \hi\ 21\,cm emission line data
  \citep{WK01} toward Mrk~876 and \pgb. Galactic blends have a
  two-component structure in these directions. High-velocity highly
  ionized gas is detected in \os\ along each sight line, but not in \nf.} 
\end{figure} 

\begin{figure}[ht]
\figurenum{4}
\epsscale{0.85}
\plotone{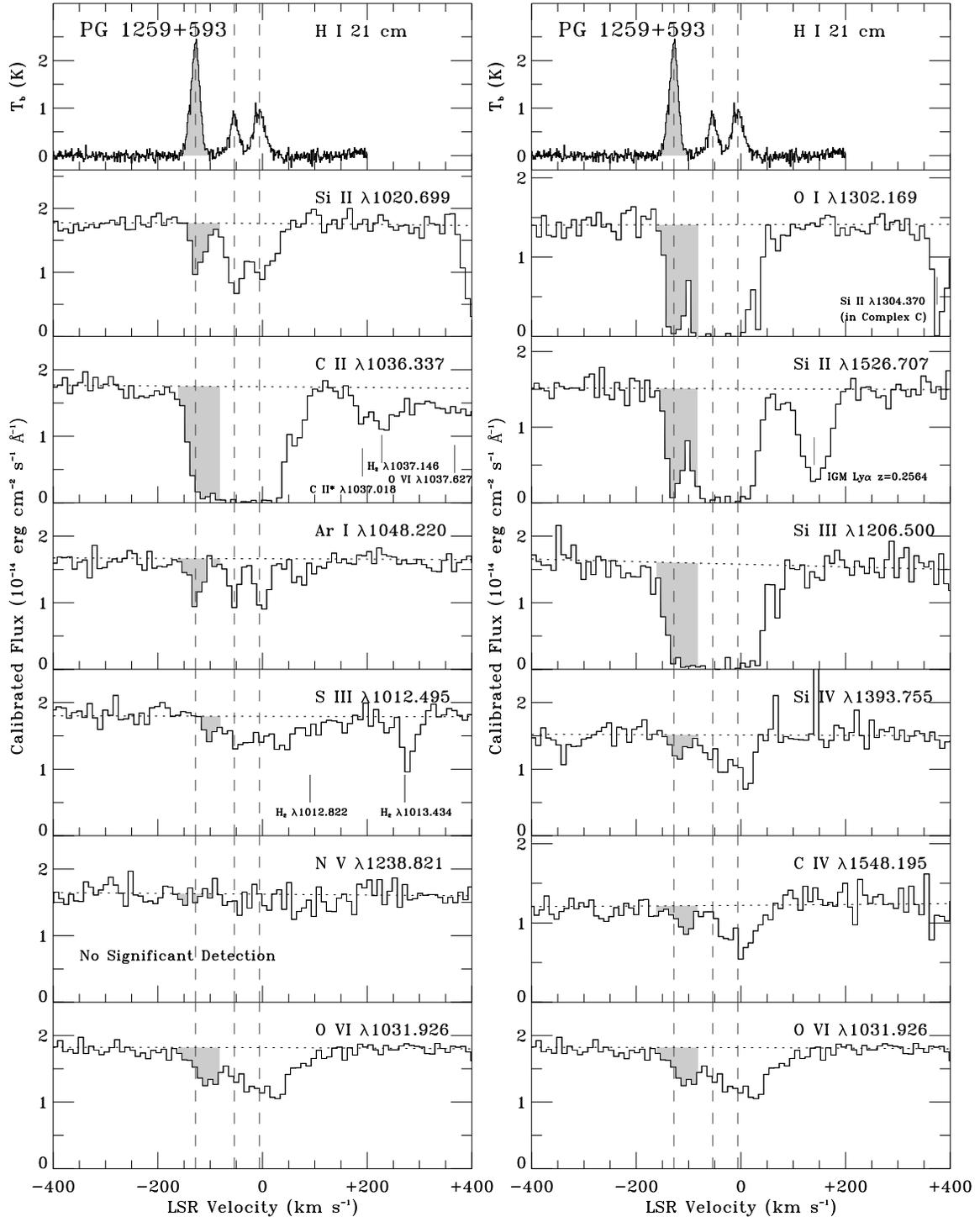}
\caption{\small \fuse, STIS, and Effelsberg 100m radio telescope
  observations of the \pga\ sight line. The \hi\ and \os\ profiles have
  intentionally been included in both columns for comparison. Highly
  ionized (\sif, \cf, and
  \os) and intermediate ionization (\ion{S}{3} and \ion{Si}{3}) gas is
  detected at Complex~C velocities ($-160<v_{LSR}<-80$\kms).}
\end{figure}

\begin{figure}[ht]
\figurenum{5}
\epsscale{0.85}
\plotone{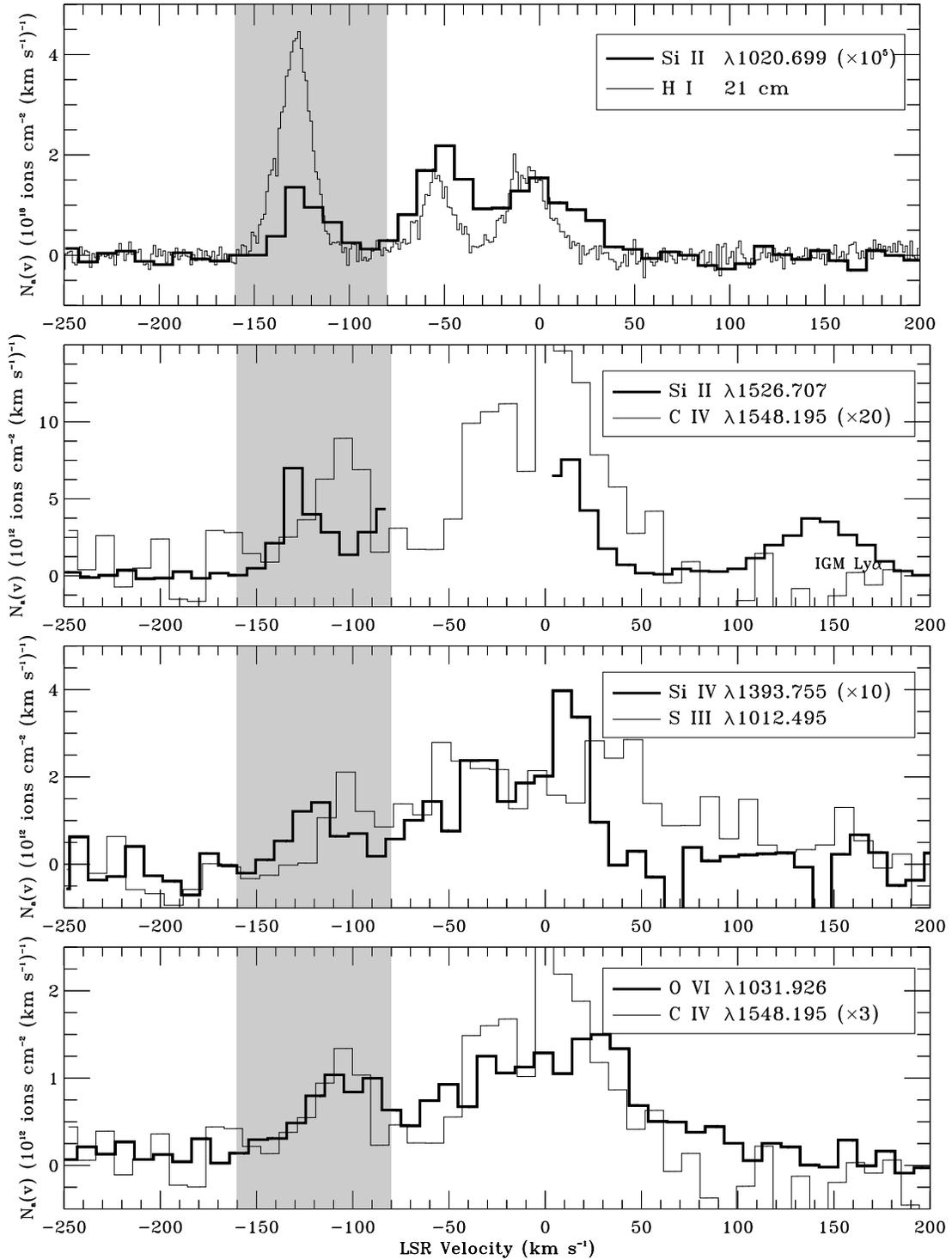}
\caption{\small Plots of apparent column density as a function of
  velocity, for various absorption lines and \hi\ emission toward
  \pga. In each panel, one of the profiles has been scaled for ease of
  comparison to the other species. The velocity range corresponding 
  to Complex~C absorption is shaded in gray. Note that in the second
  panel, we have not included \ion{Si}{2} $\lambda1526.707$ absorption
  between $v=-75$ and 0\kms, since the line is heavily saturated in
  this region. The two-component high-velocity structure is most
  easily seen in this second panel.}
\end{figure}

\begin{figure}[ht]
\figurenum{6}
\epsscale{0.85}
\plotone{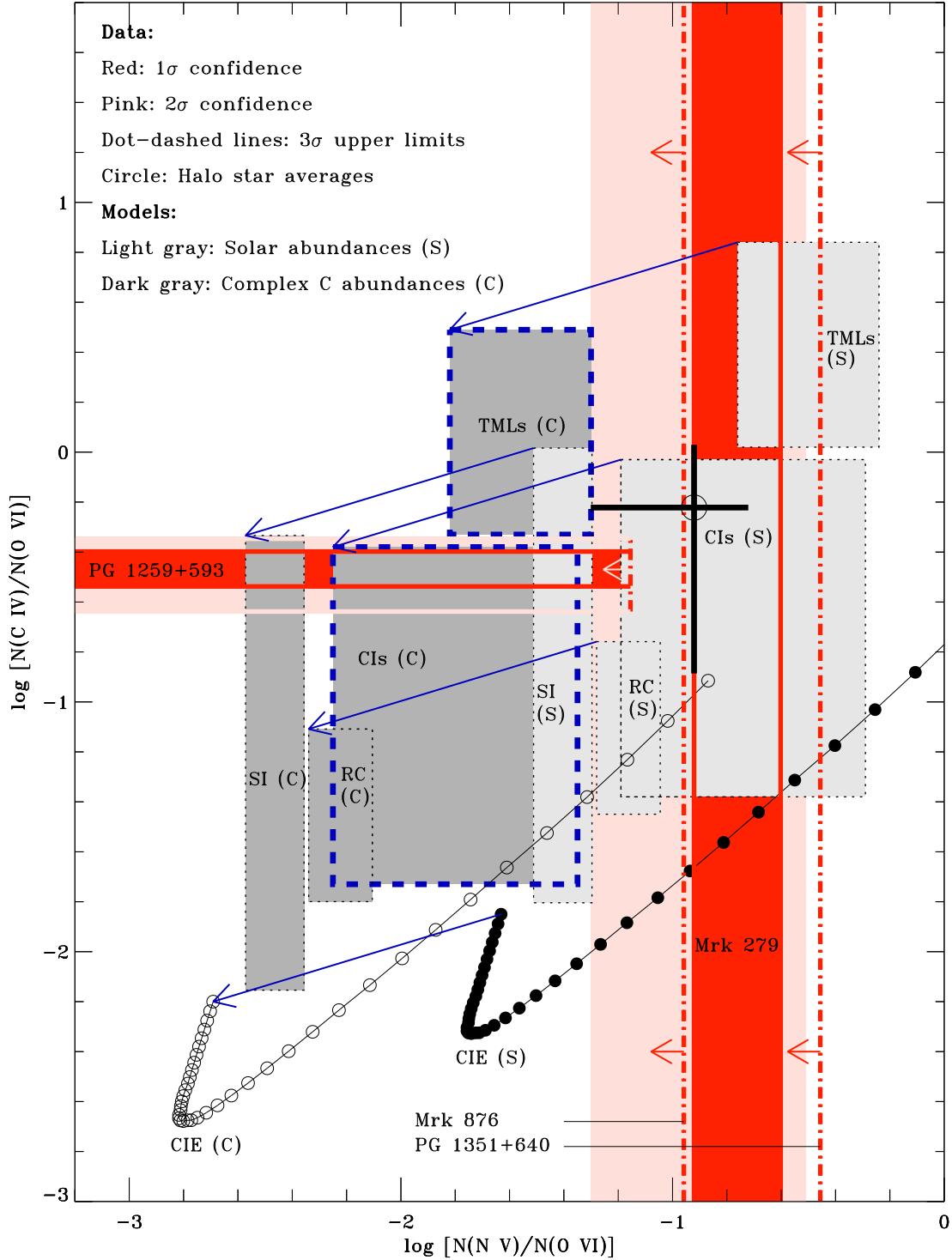}
\caption{\small Plot of $N$(\cf)/$N$(\os) against $N$(\nf)/$N$(\os)
  parameter space, comparing theory against observations. Our Mrk~279
  and \pga\ measurements are shown as colored regions, and the
  dot-dashed lines with arrows attached represent upper  
  limits to the $N$(\nf)/$N$(\os) ratio for directions where no absolute
  $N$(\nf) measurement could be made. The predictions of various
  models in two cases are shown: solar relative abundance ratios
  (light shaded regions), and assumed Complex~C 
  relative abundance ratios (dark shaded regions). The blue arrows
  connect the model predictions and show the size of the abundance
  correction we applied. Note that CIE = collisional ionization equilibrium, 
  RC = radiative cooling , CI = conductive interfaces, TML = turbulent
  mixing layers, and SI = shock ionization. The model
  parameters are described in the text.}
\end{figure}

\begin{figure}[ht]
\figurenum{7}
\epsscale{0.85}
\plotone{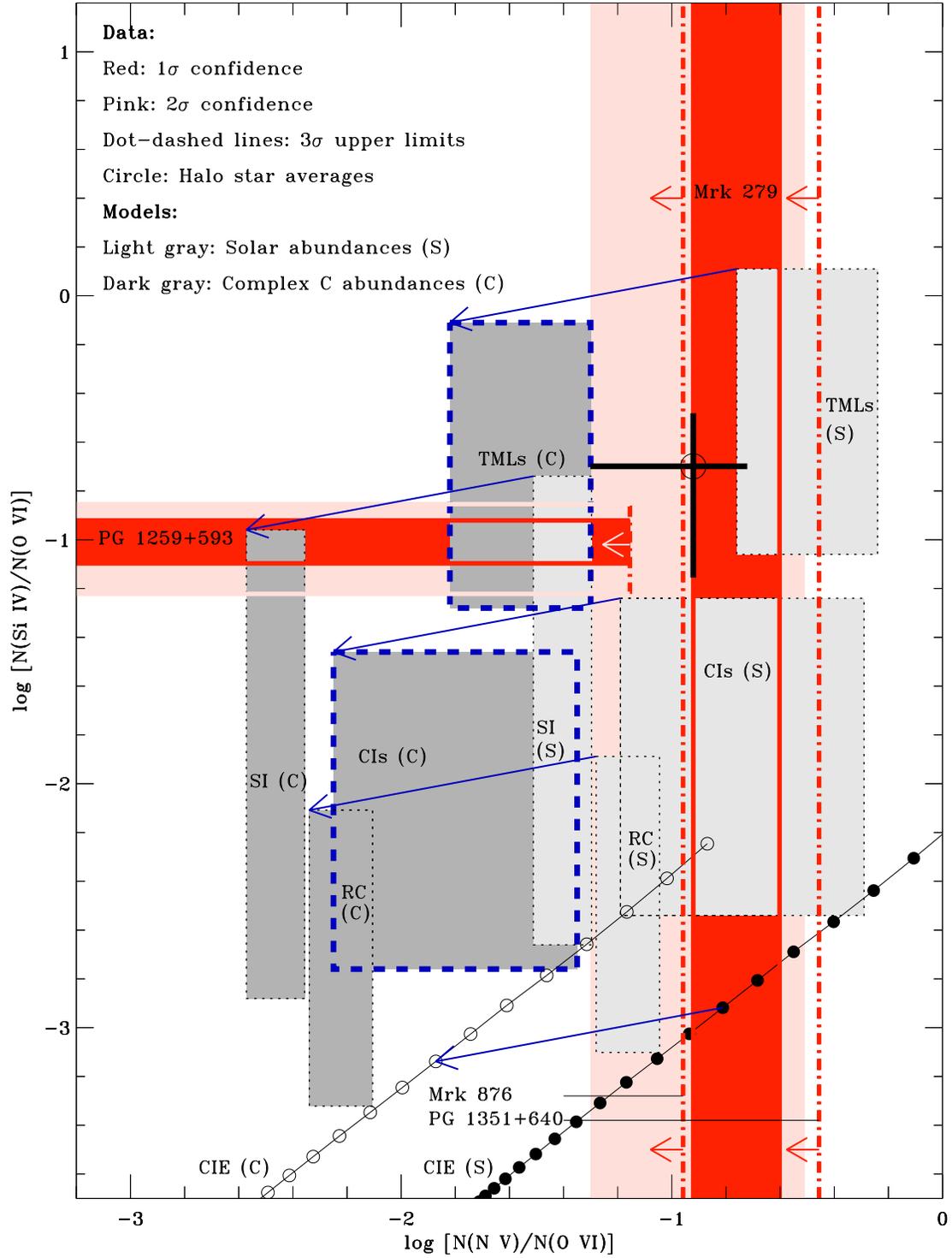}
\caption{\small Same as Figure 6, except with $N$(\sif)/$N$(\os) as
  the ordinate.}
\end{figure}

\end{document}